\documentclass{article}
\usepackage{amscd,verbatim}
\usepackage[all]{xy}
\usepackage{graphicx}

\usepackage{amsmath}
\usepackage[psamsfonts]{amssymb}
\usepackage{amsthm}
\usepackage{euscript}

\usepackage{xypic}
\input{xypic}

\usepackage{epsfig}
\usepackage{amssymb,amsmath,amsthm,amscd}

\usepackage{latexsym}

\renewcommand{\(}{\begin{equation}}
\renewcommand{\)}{end{equation} \vspace{-.05in}\linebreak}

\newcounter{saveeqn}
\newcounter{savealpheqn}

\newcommand{\alpheqn}{\setcounter{saveeqn}{\value{equation}}%
  \stepcounter{saveeqn}\setcounter{equation}{0}%
  \renewcommand{\theequation}{\mbox{\arabic{section}.\arabic{saveeqn}
\alph{equation}}}
  \renewcommand{\)}{\end{equation}}}
\def\part#1{\frac{\partial}{\partial{#1}}}%
\def\group#1{\refstepcounter{equation}\setcounter{saveeqn}{\value{equati
on}}%
  \label{#1}\setcounter{equation}{0}%
\renewcommand{\theequation}{\mbox{\arabic{section}.\arabic{saveeqn}
\alph{equation}}}
  \renewcommand{\)}{\end{equation}}}
\newcommand{\reseteqn}{\setcounter{equation}{\value{saveeqn}}%
  \renewcommand{\theequation}{\arabic{section}.\arabic{equation}}%
  \renewcommand{\)}{\end{equation}}}

\newcommand{\aalpheqn}{\setcounter{saveeqn}{\value{equation}}%
  \stepcounter{saveeqn}\setcounter{equation}{0}%
  \renewcommand{\theequation}{\mbox{
        \Alph{subsection}.\arabic{saveeqn}\alph{equation}}}
   \renewcommand{\)}{\end{equation}}}
\newcommand{\areseteqn}{\setcounter{equation}{\value{saveeqn}}%
  \renewcommand{\theequation}{\Alph{subsection}.\arabic{equation}}%
  \renewcommand{\)}{\end{equation}}}

\renewcommand{\thefootnote}{\alph{footnote}}
\renewcommand{\(}{\begin{equation}}
\renewcommand{\)}{\end{equation}}
\newcommand{\ba}{\begin{eqnarray}}
\newcommand{\ea}{\end{eqnarray}}

\newcommand{\bp}{\mathop{\vtop{\ialign{##\crcr
   $\hfil\displaystyle{}\hfil$\crcr\noalign{\kern-13pt\nointerlineskip}
   \BIG{(}\hskip0pt\crcr\noalign{\kern3pt}}}}}
\newcommand{\cbp}{\mathop{\vtop{\ialign{##\crcr
   $\hfil\displaystyle{}\hfil$\crcr\noalign{\kern-13pt\nointerlineskip}
   \BIG{)}\hskip0pt\crcr\noalign{\kern3pt}}}}}
\newcommand{\pa}{\mathop{\vtop{\ialign{##\crcr

$\hfil\displaystyle{\oplus}\hfil$\crcr\noalign{\kern+1pt\nointerlineskip
}
   \hspace{.08in}$^{\alpha=0}$\hskip6pt\crcr\noalign{\kern3pt}}}}}

\newcommand{\w}{\omega}
\newcommand{\cP}{\ensuremath{\mathcal P}}

\def\H{\ensuremath{{\mathbb H}}}

\newcommand{\beq}{\begin{equation}}
\newcommand{\eeq}{\end{equation}}

\newcommand{\iso}{\cong}

\numberwithin{equation}{section}

\catcode`\@=11
\def\vereq#1#2{\lower3pt\vbox{\baselineskip1.5pt \lineskip1.5pt
\ialign{$\m@th#1\hfill##\hfil$\crcr#2\crcr\sim\crcr}}}
\catcode`\@=12

\makeatletter
\newcommand\figcaption{\def\@captype{figure}\caption}
\newcommand\tabcaption{\def\@captype{table}\caption}
\makeatother
\renewcommand{\(}{\begin{equation}}
\renewcommand{\)}{\end{equation}}

\newcommand{\bea}{\begin{eqnarray}}
\newcommand{\eea}{\end{eqnarray}}

\DeclareMathOperator{\tmf}{tmf}

\newcommand{\C}{{\mathbb C}}
\newcommand{\R}{{\mathbb R}}
\newcommand{\Z}{{\mathbb Z}}
\newcommand{\Q}{{\mathbb Q}}

\theoremstyle{plain}
\newtheorem{theorem}{Theorem}[section]
\newtheorem{lemma}[theorem]{Lemma}
\newtheorem{prop}[theorem]{Proposition}

\theoremstyle{definition}

\begin{document}

\begin{titlepage}
%\begin{flushright}

%hep-th/yymmxxx
%\end{flushright}

\vspace{2em}
\def\thefootnote{\fnsymbol{footnote}}

\begin{center}
{\Large\bf ${\mathbb O}P^2$ bundles in M-theory} 
\end{center}
\vspace{1em}

\begin{center}
\Large  Hisham Sati
\footnote{E-mail:
{\tt hisham.sati@yale.edu}}
\end{center}

\begin{center}
\vspace{1em}
{\em { Department of Mathematics\\
Yale University\\
New Haven, CT 06511\\
USA}}\\
\end{center}

\begin{center}
Hausdorff Research Institute for Mathematics,\\
Poppelsdorfer Allee 45\\
D-53115 Bonn\\
Germany
\end{center}

\vspace{0.5cm}
\begin{abstract}
Ramond has observed that the massless multiplet of eleven-dimensional 
supergravity can be generated from the
decomposition of certain representation of the exceptional Lie group $F_4$ into those 
of its 
maximal compact subgroup 
${\rm Spin}(9)$. The possibility of a topological
origin for this observation is investigated by studying Cayley plane, ${\mathbb O} P^2$, bundles 
over eleven-manifolds $Y^{11}$. 
%Consequently, 
%the origin of the massless fields and their supersymmetry in M-theory
%are characterized geometrically and topologically.
The lift of the topological terms 
gives constraints on the cohomology of $Y^{11}$ which are derived.
%The effect of the construction on the partition function and the 
%compatibility with other physical theories is discussed.
% The various genera of ${\mathbb O} P^2$ are calculated and 
Topological  
structures and genera on $Y^{11}$
are related to corresponding ones on the total space $M^{27}$. The latter, being 27-dimensional, might 
 provide a candidate for `bosonic M-theory'. The discussion leads to 
 a connection with an octonionic version of Kreck-Stolz
 elliptic homology theory.

\end{abstract}

\vfill

\end{titlepage}
\setcounter{footnote}{0}
\renewcommand{\thefootnote}{\arabic{footnote}}

\pagebreak

\tableofcontents

\renewcommand{\thepage}{\arabic{page}}

%%%%%%%%%%%%%%%%%%%%%%%%%%%%%%%%
\section{Introduction}
%%%%%%%%%%%%%%%%%%%%%%%%%%%%%%%%

The relation between M-theory and type IIA string
theory leads to very interesting connections to K-theory \cite{DMW,
DFM} and twisted K-theory \cite{TDMW} \cite{BM1} \cite{BM2}. Exceptional groups
have also long appeared in physics. In particular, the topological 
piece of the M-theory action is encoded in part by an $E_8$ gauge 
theory in eleven dimensions \cite{Flux}. This captures the cohomology
of the $C$-field. Models
for the M-theory $C$-field were proposed in \cite{DFM} with and
without using $E_8$. The $E_8$ bundle leads to a loop bundle on the
type IIA base of the circle bundle \cite{AE} \cite{TDMW}. The role of $E_8$ and
$LE_8$ was emphasized in \cite{Sgerbe, Sloop}. In particular, in
\cite{Sgerbe} an important role for the $String$ orientation was found
within the $E_8$ construction. It is in the case when the base
$X^{10}$ is $String$-oriented that the topological action has a
WZW-like interpretation and the degree-two component of the eta-form
\cite{TDMW} is identified with the Neveu-Schwarz $B$-field
\cite{Sgerbe}.

\vspace{3mm}
In this paper we study another side of the problem, by including the
whole eleven-dimensional supermultiplet $(g, C_3, \Psi)$, i.e. the 
metric, the $C$-field, and the Rarita-Schwinger field, and not just
the $C$-field. This turns out to be related to another exceptional
Lie group, namely $F_4$, the exceptional Lie group of rank 4. 
Ramond \cite{Ram1} \cite{Ram2} \cite{Ram3} 
gave evidence for $F_4$ coming from the following two related
observations:

\begin{enumerate}
\item  $F_4$ appears explicitly  \cite{Ram3} in the light-cone formulation of 
supergravity in eleven dimensions \cite{CJS}. 
The generators $T^{\mu \nu}$ of the little group $S{\rm O}(9)$ of the Poincar\'e
group ${\rm I}S{\rm O}(1,10)$ in eleven dimensions and the spinor 
generators $T^a$ combine to form the 52 operators that 
generate the exceptional Lie algebra $\frak{f}_4$ such that 
the constants $f^{\mu \nu ab}$ in the commutation relation
\(
[T^{\mu \nu}, T^a]= if^{\mu \nu ab} T^b
\label{tij}
\)
are the structure constants of $\frak{f}_4$.
The 36 generators $T^{\mu \nu}$ are in the adjoint of $S{\rm O}(9)$ and
the 16 $T^a$ generate its spinor representation. This can be viewed as
the analog of the construction of $E_8$ out of the 
generators of $S{\rm O}(16)$ and of $E_8/S{\rm O}(16)$ in \cite{GSW}.

 \item The identity representation of $F_4$, i.e. the one corresponding
 to Dynkin index $[0,0,0,0]$, generates
the three representations of ${\rm Spin}(9)$ \cite{Ram1} 
\( 
{\rm Id}(F_4) \longrightarrow (44, 128, 84)\;,
\) 
the numbers on the right hand side correctly matching the 
number of degrees of freedoms of 
the massless bosonic content of eleven-dimensional supergravity with the individual
summands corresponding, respectively, to the graviton, the
gravitino, and the $C$-field.
% (see the beginning of section \ref{count}).
  \end{enumerate}

\vspace{3mm}
The purpose of this paper is to expand on Ramond's observations
by investigating the possibility of having an actual ${\mathbb O} P^2=F_4/{\rm Spin}(9)$
bundle over $Y^{11}$ through which the above observations can be
explained geometrically and topologically. Since $F_4$ is the isometry
group of the Cayley plane, the ${\mathbb O} P^2$ bundle will 
be the bundle associated to a principal $F_4$ bundle. We analyze some
conditions under which this is possible. 

\vspace{3mm}
In physics, the lifting of M-theory via the sixteen-dimensional manifold
${\mathbb O} P^2$ brings us to 27 dimensions. Given a Kaluza-Klein interpretation, this 
suggests the existence of a
theory in 27 dimensions, whose dimensional reduction over ${\mathbb O} P^2$
leads to M-theory. The higher dimensional theory involves spinors, and it is natural
to ask whether or not the theory can be supersymmetric. In one form
we propose this as a candidate for the `bosonic M-theory'
sought after in \cite{HS}, from gravitational geometric arguments,
and in \cite{Rey}, from matrix model arguments.

\vspace{3mm}
We consider the point of view of eleven-dimensional 
manifolds in M-theory with extra topological structure,
such as a {\it String} structure. Since any $Y^{11}$ with a {\it String} structure is zero bordant in
the {\it String} bordism group $\Omega_{11}^{\langle 8 \rangle}$ then 
this raises the question of whether there is an equivalence with a total
space of a bundle in which $Y^{11}$ is a base. For the 
Spin case, Kreck and Stolz \cite{KS} constructed an 
elliptic homology theory in which a spin manifold
of dimension $4k$ is {\it Spin} bordant to the total space 
of an $\mathbb{H} P^2$ bundle over a
zero-bordant base if and only if its elliptic genus 
$\Phi_{\rm ell} \in \mathbb{Q}[\delta, \varepsilon]$ vanishes,
where the generators $\delta$, $\varepsilon$ have degree 
4 and 8, respectively. The same authors also 
expected the existence of a homology theory based on ${\mathbb O} P^2$ 
bundles for the {\it String} case, i.e. for manifolds such that $\frac{1}{2}p_1=0$,
where $p_1$ is the first Pontrjagin class. 
So in our case, we ask whether 
there is a manifold $M^{27}$ which is an ${\mathbb O} P^2$ 
bundle over a zero bordant base and what consequence 
that has on the elliptic and the Witten genus.

\vspace{3mm} 
Some aspects of the connection to this putative homology theory are
\begin{enumerate}
\item The elliptic homology theory requires the fundamental class
\footnote{viewed as a generator.} $[{\mathbb O} P^2]$ of ${\mathbb O} P^2$ to be
inverted. This suggests connecting the lower-dimensional theory, in
our case eleven-dimensional M-theory, to a higher dimensional one
obtained by increasing the dimension by 16.

\item Previous works have used elliptic cohomology. We emphasize that
in this paper we make use of a {\it homology} theory. Thus this not
only provides further evidence for the relation between elliptic
(co)homology and string/M-theory, but it also provides a new angle
on such a relationship.

\end{enumerate}

\vspace{3mm}
In previous work \cite{KS1} \cite{KS2} \cite{KS3} \cite{S4} \cite{Stwist} evidence 
from various angles for a connection between string theory and
elliptic cohomology was given. These papers relied
heavily on analogies with the case in string theory, and were thus
not intrinsically M-theoretic.  In \cite{S1} \cite{S2} \cite{S3} a program 
was initiated to make the relation directly with M-theory. 
Thus,
from another angle, the
general purpose of this paper is two-fold:\\

$\bullet$ to point out further connections between elliptic
cohomology and M-theory

$\bullet$ to make the connection more M-theoretic, i.e. without 
reliance on any arguments from string theory.

\vspace{3mm}
${\mathbb O} P^2$ is the Cayley, or octonionic projective, plane. For an extensive 
description see \cite{Compact} \cite{Har} \cite{Baez}.
The group $F_4$ acts transitively on ${\mathbb O} P^2$, from which is follows
that ${\mathbb O} P^2 \iso F_4/{\rm Spin}(9)$. In fact $F_4$ is the isometry
group of ${\mathbb O} P^2$. The tangent space to ${\mathbb O} P^2$ at a point is
the coset of the corresponding Lie algebras $\frak{f}_4/\frak{so}(9)$,
which is ${\mathbb O}^2 \iso \R^{16}$. 

%\vspace{3mm}
%We use the Lorentz signature in studying the spectrum in section \ref{count},
%and then resort to the Euclidean signature when discussing the geometric and
%topological aspects in the rest of the paper. 

%%%%%%%%%%%%
\section{The Fields in M-theory and $\mathbb{O}P^2$ Bundles}
%%%%%%%%%%%
\label{count}
The low energy limit of M-theory (cf. \cite{Dynamics} \cite{Town}
\cite{Duff2}) is eleven-dimensional supergravity \cite{CJS}, whose 
field content on an eleven-dimensional spin manifold $Y^{11}$ with Spin 
bundle $SY^{11}$ is 
\begin{itemize}
\item {\underline{Two bosonic fields}}: The metric $g$ and the three-form $C_3$. It is 
often convenient to work with Cartan's moving frame formalism so that the metric is 
replaced by the 11-bein $e_M^A$ such that $e_M^A e_N^B= g_{MN} \eta^{AB}$,
where $\eta$ is the flat metric on the tangent space.
\item {\underline{One fermionic field}}: The Rarita-Schwinger vector-spinor $\Psi_{1}$,
which is classically a section of $SY^{11} \otimes TY^{11}$, i.e. a spinor 
coupled to the tangent bundle. 
\end{itemize}

 \vspace{3mm}
 We now give the main theme around which this paper is centered.

\vspace{3mm}
\noindent {\bf Main Idea:} {\it We interpret Ramond's triplets as arising from 
${\mathbb O} P^2$ bundles with structure group $F_4$ over our 
eleven-dimensional manifold $Y^{11}$, on which M-theory is 
`defined'.}

\vspace{3mm}
One major advantage of the introduction of an ${\mathbb O} P^2$ bundle is that 
in this picture the bosonic fields of M-theory, namely the metric and the $C$-field, can
be unified. 

\begin{theorem}
The metric and the $C$-fields are orthogonal components of the
positive spinor bundle of ${\mathbb O} P^2$. 
\end{theorem}

\proof   
The spinor bundle $S^+({\mathbb O} P^2)$ of the Cayley plane is isomorphic to
\(
S^{+}({\mathbb O} P^2) = S_0^2(V^9) \oplus \Lambda^3 (V^9),
\label{s+}
\)
where $V^9$ is a nine-dimensional vector space and 
$S_0^2$ denotes the space of traceless symmetric 2-tensors. 
This follows from an application of proposition 3 in \cite{Fried} 
which requires the study the 16-dimensional spin representations 
$\Delta_{16}^{\pm}$ as ${\rm Spin}(9)$-representations. 
The element $e_1\cdots e_{16}$ belongs to the subgroup 
${\widetilde{\rm Spin}}(9) \subset {\rm Spin}(16)$
and acts on
$\Delta_{16}^{\pm}$
 by multiplication by $(\pm1)$. 
This means that  $\Delta_{16}^{+}$ is an $SO(9)$-representation, 
but $\Delta_{16}^{-}$ is a ${\rm Spin}(9)$-representation \cite{Adams}.
Both representations do not
contain non-trivial ${\rm Spin}(9)$-invariant elements. Such 
an element would define a parallel spinor on 
$F_4/{\rm Spin}(9)$ but, since the Ricci tensor of ${\mathbb O} P^2$ is not zero, 
%(see section \ref{geom conseq})
the spinor must vanish by the Lichnerowicz formula 
\cite{Li}
$D^2 = \nabla^2 + \frac{1}{4} R_{\rm scal}$.
Then $\Delta_{16}^{+}$ as a ${\rm Spin}(9)$ representation is given by 
equation (\ref{s+}), and $\Delta_{16}^{-}$
is the unique irreducible Spin(9)-representation of dimension 128.
\endproof

%\vspace{3mm}
%\noindent {\bf Remarks}

%
%\noindent {\bf 1.} From the above we see that the Rarita-Schwinger field is given by the negative spinor bundle 
%of ${\mathbb O} P^2$. 

%\noindent {\bf 2.} The 11-bein can also be seen from the nine-dimensional bundle in another
%way. It is an element of $SL(9)/{\rm Spin}(9)$, which indeed has dimension $44$.

%\noindent {\bf 3.} In \cite{KNS} it was shown that the bosonic degrees of freedom, $g$ and $C$,
%can be assembled into an $E_{8(+8)}$-valued vielbein in eleven dimensions. As
%$E_{8(+8)}$ is the global symmetry of the two factors in the symmetry group 
%$E_{8(+8)} \times {\rm S}O(16)$,
%it would be interesting to see whether the discussion of the second factor here might 
%be related to \cite{KNS}.  

\vspace{3mm}
Thus we have

\begin{theorem}
The massless fields of M-theory are encoded in the spinor bundle of ${\mathbb O} P^2$. 
\end{theorem}

\subsection{${\mathbb O} P^2$ Bundles}
%%%%%%%%%%%%%%
\label{bund}

Having motivated ${\mathbb O} P^2$ bundles in M-theory, we now carry on with 
our proposal and construct such bundles in eleven dimensions. We
study the properties of the ${\mathbb O} P^2$ bundle as well as the 
associated $F_4$ bundle and give some consistency conditions.
As bundles are characterized by characteristic classes and genera,
we `compare' the structure of the base and that of the total space.
For that purpose we start with discussing the relevant genera of the 
fiber. 

%%%%%%%
\subsection{Genera of ${\mathbb O} P^2$}
%%%%%%%%%
\label{genera}

A genus is a function on the cobordism ring  $\Omega$ (see section \ref{top} for
cobordism). More precisely, it is
a ring homomorphism $\varphi :  \Omega \otimes R \to R$, where $R$ is any
integral domain over $\Q$. It could be $\Z$, $\Z_p$ or $\Q$ itself. Genera in general
have expressions given in terms of characteristic classes. Two important 
`modern'
genera are
the elliptic genus $\Phi_{\rm ell}$ and the Witten genus $\Phi_{W}$. 
The first is characterized by two parameters,
denoted $\varepsilon$ and $\delta$, whose various values give different specializations 
of $\Phi_{\rm ell}$. 
Special values of the parameters correspond to more `classical' genera. 
The values $\delta=\varepsilon=1$ leads to the $L$-genus
$L : \Omega \otimes \Q \to \Q$, and the values $\delta=-\frac{1}{8}$, $\varepsilon=0$
leads to the $\widehat{A}$-genus $\widehat{A}: \Omega \otimes \Q \to \Q$. Depending 
on the type of cobordism considered, $\Omega$ and also $R$ can vary.
For instance, when the manifolds are Spin then the $\widehat{A}$-genus is 
an integer and so $\widehat{A} : \Omega^{\rm Spin} \otimes \Z \to \Z$. 
The Witten genus is defined for any topological manifold but it becomes a modular
form for
special manifolds, namely ones with a {\it String}
structure or $BO\langle 8 \rangle$-structure, and those are the manifolds that
satisfy $\frac{1}{2}p_1=0$, where $p_1$ is the first Pontrjagin class of the
tangent bundle. The Witten genus is a map
$\Phi_W : \Omega^{BO\langle 8 \rangle} \otimes R \to 
MF=R[E_4, E_6]$, where $MF$ is the ring of modular forms generated by the
Eisenstein series $E_4$ and $E_6$, and $R$ is usually $\Q$ or $\Z$. We describe
this more precisely below.  

\vspace{3mm}
It is natural to ask what the values of the elliptic genus and of the 
Witten genus of ${\mathbb O} P^2$ 
are. First, however, we consider the classical genera.

\paragraph{1. The classical genera.}

We give the following specialization.

\begin{lemma}
\begin{enumerate}
\item The ${\widehat A}$-genus of ${\mathbb O} P^2$ is zero, $\widehat{A}[{\mathbb O} P^2]=0$.
\item The $L$-genus of ${\mathbb O} P^2$ is $u^2$, where $u$ is the generator of
$H^8({\mathbb O} P^2; \Z)$. 
\end{enumerate}
\label{AL}
\end{lemma}

\paragraph{2. The Witten genus.}

Next we consider another genus, the Witten genus, which
 can be defined in the following way. 
There is a convenient collection of manifolds $\{M^{4n}\}$ that generate
the rational cobordism ring $\Omega \otimes \Q$ \cite{24}. The advantage 
of this basis is that each $M^{4n}$ has a single nonzero Pontrjagin 
class, the top one $p_n=d_n (2n-1)! m$ where $m$ generates $H^{4n}(M^{4n})$.
On this basis,
$\Phi_W (M^{4k})= {\rm num}_k E_{2k}$ for $k> 1$ and
$\Phi_W(M^4)=0$, where ${\rm num}_n/d_n=B_{2n}/4n$ is the 
given numerator,
with ${\rm num}_n$ and $d_n$ relatively prime, 
and $B_{2n}$ the even Bernoulli numbers. 
%This means that 
%we use 
%\(
%\sigma (z, \tau)=z \exp \left( - \sum_{k \geq 2}
%\frac{2}{(2k)!}
% \frac{B_{2k}}{4k}E_{2k}(\tau)
%z^{2k}\right).
%\)
%This then comes
%from a logarithm of some series $Q(z)$, which
%is found from
%\bea
%\log (Q(z))&=& \sum_{k \geq 2} 2\frac{{\rm num}_k}{d_k}
%E_{2k} \frac{z^{2k}}{(2k)!}
%\nonumber\\
%&=&
%\sum_{k \geq 2} 2 \frac{G_{2k} z^{2k}}{(2k)!}
%\nonumber\\
%&=&
%\log \left(\frac{z}{\sigma(z, \tau)}  \right),
%\eea
%which shows that $Q(z)=z/\sigma(z, \tau)$.
The ring of modular forms for the full modular group 
is (cf. \cite{AHS}) $MF=\Z [E_4, E_6, \Delta]/(E_4^3-E_6^2 - 1728\Delta)$, where
$\Delta=q \prod_n (1-q^n)^{24}$. 
By inspecting the Bernoulli numbers we can see that the first four terms in 
$d_n$ are $24, 240, 504, 480$. This is enough for working 
up until real dimension 16. 

\begin{theorem}
The Witten genus of ${\mathbb O} P^2$ is zero, $\Phi_W({\mathbb O} P^2)=0$.
\label{genus}
\end{theorem}

\proof
${\mathbb O} P^2$ has positive scalar curvature, so its ${\widehat A}$-genus is zero
$\widehat{A}({\mathbb O} P^2)=0$.  ${\mathbb O} P^2$ is also a {\it String} manifold, so its
Witten genus $\Phi_W({\mathbb O} P^2): \Omega_{16}^{BO\langle 8 \rangle}
=\pi_{16}MO\langle 8 \rangle \to \pi_{*}eo_2=MF_*$ 
must be a modular form for $SL(2,\Z)$ of weight
equals to half its dimension \cite{Zag}, i.e. 8.  What modular forms do we have?
The ring of integral modular forms is (cf. \cite{AHS})
\(
MF_*=\Z [ E_4, E_6, \Delta]/(2^6 \cdot 3^3 \Delta - E_4^3 + E_6^2)
\)
where $E_4 \in MF_4$, $E_6 \in MF_6$, and $\Delta \in MF_{12}$.
Thus the only modular form of weight 8 is $E_4^2$. 
However the form of the Eisenstein series is 
$E_4= 1+ $ higher terms, so that the required modular form does not start 
with zero. Therefore $\Phi_W({\mathbb O} P^2)=0$.  
\endproof

\paragraph{3. The elliptic genus.}

Next we consider the elliptic genus $\Phi_{\rm ell}: \Omega_{*}^{BO \langle 8 \rangle}
\otimes \Q \to \Q [\delta, \varepsilon]$, where the generators $\delta$ and $\varepsilon$
have degrees 4 and 8, respectively. 

\begin{theorem}
The elliptic genus of the Cayley plane is $\Phi_{\rm ell} ({\mathbb O} P^2)=\varepsilon^2$.
\label{ellop2}
\end{theorem}

\proof
There are several ways to prove this. The first one is to use the idea
of cobordism as in the proof of the case of the quaternionic projective plane $\mathbb{H}P^2$.
%Lemma \ref{ellhp2}. 
However, we can simply apply a result from \cite{HiS}. Since ${\mathbb O} P^2$ is a connected homogeneous
space of a compact connected Lie group $F_4$, and since ${\mathbb O} P^2$ is oriented and
admits a Spin structure, then the normalized elliptic genus 
$\Phi_{\rm norm}:=\Phi_{\rm ell}/{\varepsilon}^2$ is a constant modular function
\(
\Phi_{\rm norm}({\mathbb O} P^2) =\sigma ({\mathbb O} P^2).
\)
Thus we immediately get the result.
\endproof

%
%\vspace{3mm}
%\noindent {\bf Remark.}
%The $F_4$-equivariant elliptic genus of ${\mathbb O} P^2$, i.e. the index of the loop
%signature operator $\S$
%\(
%{\rm Index}_{F_4} {\S}= 
%L({\mathbb O} P^2) {\rm ch} \left\{ \bigotimes_{n \geq 1} S_{q^n} (T {\mathbb O} P^2) 
%\bigotimes_{m \geq 1} \Lambda_{q^m} (T {\mathbb O} P^2) 
%\right\} [{\mathbb O} P^2]\;,
%\)
%is then just the signature $\sigma ({\mathbb O} P^2)$. 

\paragraph{4. The Ochanine genus.}

We next consider the Ochanine genus \cite{Och2}, which is a generalization of the
elliptic genus in such a way that it involves $q$-expansions. The Ochanine genus
is a ring homomorphism
\(
\Phi_{\rm och} : \Omega_*^{\rm spin} \longrightarrow KO_* ({\rm pt}) [[q]],
\label{defoch}
\)
from the Spin cobordism ring to the ring of power series with coefficients 
in 
\(
KO_*({\rm pt})= \Z \left[ \eta, \omega, \mu, \mu^{-1} \right]/ \left( 
2 \eta, \eta^3, \eta \omega, \omega^2 - 2^2 \mu \right), 
\label{kopt}
\)
where $\eta \in KO_1, \omega \in KO_4$, and $\mu\in KO_8$ are 
generators of degrees 1, 4, and 8, respectively, and 
are given by the normalized Hopf bundles $\gamma_{{}_{\R P^1}}-1$,
$\gamma_{{}_{\H P^1}}-1$, $\gamma_{{}_{{\mathbb O} P^1}}-1$ 
(viewed as real vector bundles) over the real, quaternion, and Cayley 
projective lines $\R P^1=S^1$, $\H P^1=S^4$, and 
${\mathbb O} P^1=S^8$.

\vspace{3mm}
For a manifold $M^m$ of dimension $m$, corresponding to the projection
map $\pi^{M^m} : M^m \to {\rm pt}$ there is the Gysin map 
$\pi_{!}^{M^m}: KO(M^m) \to KO^m({\rm pt})=KO_m({\rm pt})$. Now consider a real 
vector bundle $E$ on $M^m$ and form the following
combination of exterior powers and symmetric powers of $E$
\(
\Theta_q(E) = \sum_{i \geq 0} \Theta^i (E) q^i 
= \bigotimes_{n \geq 1} \left( \Lambda_{-q^{2n-1}}(E) \otimes S_{q^n}(E) \right),
\)
which, since it is multiplicative under Whitney sum, can be considered as 
an exponential map
$\Theta_q : KO(M^m) \to KO(M^m) [[q]]$. Now specialize $E$ to be the 
reduced tangent bundle $\widetilde{TM^m}$, which is $TM^m -m$. Then 
the Ochanine genus is defined to be \cite{Och2} \cite{KS}
\bea
\Phi_{\rm och} (M^m) &:=& \sum_{i \geq 1} \Phi_{\rm och}^i (M^m) q^i
\nonumber\\
&=& \sum_{i \geq 0} \pi_{!}^{M^m} \left( \Theta^i( \widetilde{TM^m}) \right)
q^i
\nonumber\\
&=& \theta(q)^{-m} \langle \; \Theta_q(TM^m), [M^m]_{KO} \; \rangle 
 \in KO_m({\rm pt})[[q]],
 \label{ochdef}
\eea
where $[M^m]_{KO} \in KO_m(M^m)$ denotes the Atiyah-Bott-Shapiro
orientation \cite{ABS} of $M^m$, $\langle\; , \; \rangle: KO^i (X) \otimes KO_j(X) \to KO_{j-i}$
is the Kronecker pairing, and 
\(
\theta(q): = \Theta_q(1)=\prod_{n \geq 1} \frac{1-q^{2n-1}}{1-q^{2n}}=1 -q + q^2-2q^3 
\pm \cdots ~~~ \in \Z [[q]],
\)
is the Ochanine genus of the trivial line bundle.

\vspace{3mm}
The degree zero part $\Theta^0(E)$ is a trivial real line bundle, and
corresponds to the Atiyah invariant 
$\Phi_{\rm och}^0(M)= \pi_{!}^{M^m}(1)=
\langle 1, [M^m]_{KO} \rangle=\alpha(M^m)$.
The cobordism invariant $\alpha \in KO_m$ \cite{At} can 
be interpreted as the index of a family of operators associated to 
$M^m$ parametrized by $S^m$ \cite{Hit}. Thus the $\alpha$-invariant is the 
classical value of the Ochanine genus in the same way that the $\widehat{A}$-genus
and the $L$-genus are the classical values of the elliptic genus corresponding,
respectively, to
\bea
\delta=\widehat{A}(\C P^2)&=&-\frac{1}{8}, ~~~~~~~~~~~~~~~~~~~
\varepsilon=\widehat{A}(\H P^2)=0,~~{\rm and}
\nonumber\\
\delta=L(\C P^2)&=&0, ~~~~~~~~~~~~~~~~~~~~~~\varepsilon=L(\H P^2)=1.
\eea
The Ochanine genus is related to the restriction $\Phi_{\rm ell, int}$ 
to $\Omega_*^{\rm spin}$ of the universal
elliptic genus $\Phi_{\rm ell, uni}: \Omega_*^{SO} \to \Q[[q]]$, whose parameters are 
\bea
\delta&=&-\frac{1}{8}-3 \sum_{n \geq 1} \left( \sum_{d|n, ~d~{\rm odd}} d\right) q^n
=
-\frac{1}{8} + q-{\rm expansion},
\nonumber\\
 \varepsilon&=& \sum_{n \geq 1} \left( 
 \sum_{{ d|n,~   \frac{n}{d}~{\rm odd}}} d^3 \right)q^n
= 0 + q-{\rm expansion}.
\label{ep}
\eea
More precisely, $\Phi_{ell, int}=Ph \circ \Phi_{\rm och} : \Omega_*^{\rm spin} \to \Z[[q]]$,
where $Ph$ is the Pontrjagin character
\(
Ph: KO^*(X) \longrightarrow^{\!\!\!\!\!\!\!\!\!\!\otimes \C}~K^*(X)\longrightarrow^{\!\!\!\!\!\!\!\!\!\!\rm ch}
~~H^{**}(X;\Q), 
\)
which can be thought of as the analog for real vector bundles of the Chern character for 
complex vector bundles.

\vspace{3mm}
We now check the value of
$\Phi_{\rm och}$ for ${\mathbb O} P^2$.

\begin{theorem}
The Ochanine genus of ${\mathbb O} P^2$ is 
$\Phi_{\rm och} ({\mathbb O} P^2)= \varepsilon^2 \mu^2$.  
\label{ochop2}
\end{theorem}

\proof
The Ochanine genus $\Phi_{\rm och}({\mathbb O} P^2)$ is the map 
$\Omega_{16}^{\rm spin} \to KO_{16}[[q]]$.
Note that $\Omega_{16}^{\rm spin}=\Z \oplus \Z \oplus \Z$ and
that $KO_{16}({\rm pt})= \Z$ with 
generator $\mu^2$.
The image $\Phi_{\rm och} (\Omega_{16}^{\rm spin})$ is the set of all
modular forms of degree 16 and weight 8 over $KO_{16}=\Z$.
Let $M^{\Gamma}(KO_{16})$ be the graded ring of modular forms
over $KO_{16}$ for $\Gamma$, a subgroup of finite index in
$SL(2;\Z)$. For $M_*^{\Gamma} (\Z)=\Z[\delta_0, \varepsilon]$,
where $\delta_0=-8\delta \in M_2^{\Gamma} (\Z)$ and 
$\delta$ and $\varepsilon \in M_4^{\Gamma}(\Z)$ are 
the generators in (\ref{ep}), we have 
\bea
M^{\Gamma} (KO_{16}) &\iso& KO_{16} \otimes M_*^{\Gamma} (\Z)
\nonumber\\
&=& \Z \otimes \Z[\delta_0, \varepsilon]\; .
\eea
Then a modular form of degree 16 and weight 8 can be written in 
a unique way as a polynomial $P(\delta_0, \varepsilon)$ of weight 8 with 
integer coefficients. Still applying the construction in \cite{Och2}, the 
Ochanine genus in our case is 
\(
\Phi_{\rm och} ({\mathbb O} P^2)= \left( 
a_0({\mathbb O} P^2) \delta_0^4 + a_1({\mathbb O} P^2) \delta_0^2 \varepsilon 
+ a_2({\mathbb O} P^2) \varepsilon^2 \right) \mu^2,
\)
with uniquely defined homomorphisms , for $i=1,2,3$,
\(
a_i \cdot \mu^2 : \Omega_{16}^{\rm spin}=\Z \oplus \Z \oplus \Z 
\longrightarrow KO_{16}=\Z \; .
\)
The integers $a_i$ can be determined as follows. We have already seen that
the lowest coefficient is given by the Atiyah invariant. Since ${\mathbb O} P^2$
admits a Riemannian metric of positive scalar curvature
% (see section \ref{geom conseq}) 
then, from 
\cite{Hit}, $\alpha ({\mathbb O} P^2)=0$, and hence we have determined that
$a_0 ({\mathbb O} P^2)=0$. Another way of seeing this is to notice that for manifolds
of dimension $4k$, the Atiyah invariant is essentially the $\widehat{A}$-genus,
which, by Lichnerowicz theorem \cite{Li}, vanishes for a manifold with positive scalar
curvature. The highest coefficient,
$a_2({\mathbb O} P^2)$, is given by the Ochanine $k$-invariant, which in 
this case is just the signature
$
a_2({\mathbb O} P^2)= \sigma ({\mathbb O} P^2)=1$.
It remains to calculate $a_1$. This is given by the first KO-Pontrjagin class
$\Pi_1$  
\(
a_1 ({\mathbb O} P^2)=  \Pi_1 (T {\mathbb O} P^2) = -\Lambda^1 (T {\mathbb O} P^2- 16),
\label{a1}
\)
which is just $-(T {\mathbb O} P^2 - 16)$. The KO-Pontjagin classes are defined as follows
\cite{ABP2}. For an $n$-dimensional vector bundle $\xi$ over a space $X$, 
$\Pi_u (\xi) \in KO^0(X)$ are defined by 
\(
(1+t)^n \sum_{k=0}^{\infty}\frac{t^k}{(1+t)^{2k}}
\Pi_k(\xi)=\sum_{k=0}^{\infty} t^k \Lambda^k (\xi)\; .
\label{KO Pont}
\)
 For $k=1$ this gives 
the first KO-Pontrjagin class used in (\ref{a1}).
Alternatively, we can look at the $q$-components of $\Phi_{\rm och}$ from the 
first line of equation (\ref{ochdef}) and get
\bea
\Phi_{\rm och}^0 ({\mathbb O} P^2) &=& \langle 1\; ,\; [{\mathbb O} P^2]_{KO} \rangle= \alpha ({\mathbb O} P^2) 
\nonumber\\
\Phi_{\rm och}^1 ({\mathbb O} P^2) &=& \langle - \Pi_1({\mathbb O} P^2)\; ,\; [{\mathbb O} P^2]_{KO} \rangle
= \langle - ( T {\mathbb O} P^2 - 16) \; , \; [({\mathbb O} P^2)]_{KO} \rangle.
\eea
We still have to calculate $a_1$. We use the topological Riemann-Roch
theorem (see \cite{StongNotes}) which states that for $M$ a closed Spin manifold
and $x \in \widetilde{KO}^*(M)$, then 
$Ph \langle x , [M]_{KO} \rangle = \langle \widehat{A}(M) Ph (x), [M] \rangle_H$,
where $\langle~,~\rangle_H$ is the Kronecker pairing on cohomology.
Taking $M={\mathbb O} P^2$ and $x=T{\mathbb O} P^2$, we get for $a_1$  
\(
\langle\; \widehat{A}({\mathbb O} P^2) Ph (T {\mathbb O} P^2)\; , \; [{\mathbb O} P^2]\; \rangle_H,
\)
which is zero because, as we have seen,  $\widehat{A}({\mathbb O} P^2)=0$.
\endproof

%
%\begin{corollary}
%The $q$-expansion of $\Phi_{\rm och} ({\mathbb O} P^2)$ is 
%$\mu^2 q^2 (1 + 16q + 120 q^2 + 576 q^3 + \cdots)$. 
%\label{corq}
%\end{corollary}
%\proof
%We expand $\varepsilon$ from (\ref{ep}) to get
%\(
%{\varepsilon} = q + 8q^2 + 28q^3 + 64q^4 + 126q^5 + 224q^6 + 344q^7 + \cdots,
%\)
%so that
%\(
% {{\varepsilon}}^2 = q^2 + 16q^3 + 120q^4 + 576q^5 + \cdots .
%\)
%\endproof

%\noindent {\bf Remarks}

%
%\noindent {\bf 1.} The $\widehat{A}$-genus is obtained  from the Ochanine genus by setting
%$\varepsilon$ to zero, or equivalently setting $q$ to zero in our case. We can see from
%the above expressions in either Theorem \ref{ochop2} or from Corollary \ref{corq}
%that indeed we do reproduce $\widehat{A}({\mathbb O} P^2)=0$.

%\noindent {\bf 2.} The signature is obtained from the Ochanine genus by replacing 
%$\delta$ by $-1$, $\varepsilon$ by 1, $\omega$ by 2, and $\mu$ by 1. 
%Applying these transformations to $\Phi_{\rm och}({\mathbb O} P^2)$ gives 1,
%which is indeed $\sigma({\mathbb O} P^2)$, cf. (\ref{sign}).

%

%%%%%%%%%%%%
\subsection{ ${\mathbb O} P^2$ bundles over eleven-manifolds}
%%%%%%%%%%%%

Consider the fiber bundle $E \to Y^{11}$ with fiber ${\mathbb O} P^2$ and
structure group $F_4$. There is a universal bundle of this type.
 ${\mathbb O} P^2$ bundles over $Y^{11}$ are 
pullbacks of the universal bundle 
\( 
{\mathbb O} P^2=F_4/{\rm Spin}(9) \longrightarrow
B{\rm Spin}(9) \longrightarrow BF_4 
\label{main} 
\) 
by the classifying map $f :
Y^{11} \to BF_4$.
 In this paper we will consider the diagram
 \(
  \xymatrix{
  {\mathbb O} P^2
  \ar[rr]
%\ar[rrdd]
  &&
   M^{27}
   \ar[rrdd]
   \ar[dd]_{\pi}
   &&
  % \ar[dd]^{\times 8}
  &&
   \\
   \\
   &&
   Y^{11}
    \ar[rr]_{f}
  &&
BF_4\;.
  }
  \label{diag}
  \)
Note that the map from $M^{27}$ to $BF_4$ can be $f \pi$
and this will be useful later in section \ref{top}.
We first have the following.

\begin{prop}
The obstruction to existence of a section of an ${\mathbb O} P^2$ fiber bundle 
over an eleven-dimensional manifold $Y^{11}$ lies in 
$H^9(Y^{11}; \Z)$, $H^{10}(Y^{11}; \Z_2)$ and $H^{11}(Y^{11}; \Z_2)$.
\end{prop}

\proof
For a fiber bundle $F \to E \to B$, the existence to having a section lies
in the groups $H^r\left( B; \pi_{r-1}(F)\right)$ for all nonzero $r \in \mathbb{N}$. 
In our case, ${\mathbb O} P^2$ has $\pi_i=0$ for $i \leq 7$, so that the first obstruction 
is in $H^9\left( Y^{11}; \pi_8({\mathbb O} P^2)\right)$, which is $H^9(Y^{11}; \Z)$.
The next two nontrivial homotopy groups of ${\mathbb O} P^2$, both are $\Z_2$, in 
dimension 9 and 10 so that the obstructions are in $H^{10}(Y^{11}; \Z_2)$
and $H^{11}(Y^{11}; \Z_2)$. ${\mathbb O} P^2$ has further nontrivial homotopy groups
but that would bring us to $H^{\geq 12}$, which are zero for an eleven-manifold.
\endproof

\noindent {\bf Remarks}

\noindent {\bf 1.} The first obstruction $H^9(Y^{11};\Z)$ is called the primary 
obstruction. 

\noindent {\bf 2.} Note that the primary obstruction is a $\Z$-class whereas the
secondary obstructions are $\Z_2$-classes.
 
\vspace{3mm}
In forming bundles with ${\mathbb O} P^2$ as fibers, we are forming bundles of 
$BO \langle 8 \rangle$-manifolds
over $Y^{11}$. We will next investigate the relation between structures on
$Y^{11}$, on the fiber ${\mathbb O} P^2$, and on the total space $M^{27}$.

%%%%%%%%%%%%%%%%%%%%%%
\subsection{Relating $Y^{11}$ and $M^{27}$}
\subsubsection{Topological consequences: the higher structures}
%%%%%%%%%%%%%%%%%%%%%%%%
\label{conseq}
We ask the question whether topological conditions on $Y^{11}$, namely
having {\it Spin}, {\it String}, or {\it Fivebrane} structure \cite{SSS1} \cite{SSS2}, will 
lead to (similar) structures
on $M^{27}$. The answer to such a question is possible because 
we know about the (non-)existence of these structures on ${\mathbb O} P^2$.

\vspace{3mm} 
The condition $\lambda:=\frac{1}{2}p_1=0$ for lifting the structure
group of the tangent bundle to ${\rm String}(n)$ is related to the
condition $W_7=0$ for orientation with respect to either the $p=2$
integral Morava K-theory $K(2)$ or Landweber's elliptic cohomology
theory $E(2)$ \cite{KS1}. The first condition implies the second,
but the converse is not true, a counterexample 
being $X^{10}=S^2 \times S^2 \times \C P^3
$ \cite{KS1}. Thus if we assume the {\it String} orientation,
then we are already guaranteed the $W_7$ orientation, and so the discussion
and constructions in \cite{KS1} \cite{KS2} \cite{KS3} \cite{S4} for ten-dimensional
string theory apply. The condition $\lambda=0$ can be extended 
from ten to eleven dimensions and vice versa. This is because
for $Y^{11}=X^{10} \times S^1$ the first Pontrjagin classes are
related as (using bundle notation) 
$p_1(TX^{10} \oplus TS^1) = p_1(TX^{10}) + p_1(TS^1)$,
but for dimensional reasons $p_1(TS^1)=0$ so that we
have $p_1(Y^{11})= p_1(X^{10})$. Thus the {\it String} condition 
can be translated from M-theory to string theory and back as
desired. 

\vspace{3mm}
There is no cohomology in degree four for ${\mathbb O} P^2$, so we immediately have

\begin{prop}
${\mathbb O} P^2$ admits a $BO \langle 8 \rangle$-structure.
\end{prop}

\noindent {\bf Remark.}
If $Y^{11}$ is a $BO\langle 8 \rangle$-manifold, i.e. is
$MO\langle 8 \rangle$-orientable, then it has an $MO\langle 8
\rangle$ homology fundamental class, \( [Y^{11}]_{MO\langle
8\rangle} \in MO\langle 8 \rangle_{11}(Y^{11}). \) Any integral
expression will involve this class. This would also enter the
construction of the $BO\langle 8 \rangle$ partition function.

\vspace{3mm}
We would like to check to what extent we can know the cohomology of
the total space $M^{27}$ in terms of the cohomology of the base $Y^{11}$, given 
that we know the cohomology of the fiber ${\mathbb O} P^2$. One way to detect this 
is by using the Serre spectral sequence for the bundle
\(
E_2^{p, q} =H^p \left(Y^{11}, H^q({\mathbb O} P^2) \right)
\Rightarrow H^{p+q} (M^{27})\; .
\)
Consider the case of a product $M^{27} = {\mathbb O} P^2 \times Y^{11}$, i.e. when the
bundle is trivial. In this case,  using the K\"unneth theorem 
%which for a field
%$F$ (e.g. $\C$, $\R$, $\Q$, $\Z_p$) is 
%\(
%H^n(M^{27};F)= H^n ({\mathbb O} P^2 \times Y^{11};F) \cong
%\bigoplus_{i+j=n} H^i({\mathbb O} P^2; F) \otimes_{F} H^j(Y^{11};F),
%\)    
%and for a ring $R$ (e.g. $\Z$), using the universal coefficient theorem, is
%the split sequence
%\bea
%0 \longrightarrow
%\bigoplus_i \left( H^i({\mathbb O} P^2 ;  R) \otimes_R H^{n-i} (Y^{11};R) 
%\right)
%\longrightarrow
%H^n({\mathbb O} P^2 \times Y^{11} ; R) \longrightarrow
%\nonumber\\
%\bigoplus_i {\rm Tor}_R \left( H^i({\mathbb O} P^2 ; R) ; H^{n-i-1}(Y^{11} ;R) \right)
%\longrightarrow 0 \; .
%\eea
%Then we have, for any coefficients $C$,
%\(
%H^n({\mathbb O} P^2 \times Y^{11}; C) \cong
%\bigoplus_pH^p\left( Y^{11} ; H^{n-p}({\mathbb O} P^2;C) \right)\; .
%\)
and the fact that the cohomology of ${\mathbb O} P^2$ is nonzero only in degrees 8 and 
16, we get 
\begin{prop}
\(
H^n({\mathbb O} P^2 \times Y^{11} ; C) \cong
H^{n-8} (Y^{11} ; C) \oplus H^{n-16}(Y^{11}; C)\;.
\)
\end{prop}

\vspace{3mm}
We next consider the case when the bundle is not trivial. A simplification 
is made if coefficients are taken so that the cohomology of the fiber is
trivial in those coefficients. The torsion (`bad') primes for $F_4$ are 
2 and 3, so that one might expect that those are the primes that
do not cause such a simplification. It will turn out that this is true only
for $p=3$, as we now show. We first show that $p=3$ occurs and then
that it is the only one. 

\vspace{3mm}
The cohomology of the classifying spaces of 
${\rm Spin}(9)$ and $F_4$ with $\Z_p$ coefficients,
$p=2,3$, are as follows.
The cohomology ring of $BF_4$ with coefficients in $\Z_2$ is given by the
polynomial ring \cite{Borel} 
\( 
H^*( BF_4;\Z_2)=\Z_2 \left[ x_4, x_6,
x_7, x_{16}, x_{24}\right], 
\label{bfz2first}
\) 
where $x_i$ 
are polynomial generators of degree $i$ related by the Steenrod square
operation $Sq^i : H^n(BF_4 ; \Z_2) \to H^{n+i}(BF_4; \Z_2)$ as 
\(
x_6=Sq^2 x_4, ~~~~~ x_7= Sq^3 x_4, ~~~~~ x_{24}= Sq^8 x_{16}\; .
\)
$H^*(BF_4; \Z_3)$ is generated by $x_i$ for $i=$ 4, 8, 9, 20, 21, 25, 26, 36, 48,
with the structure of a polynomial algebra \cite{Toda}.
Considering $p=3$, this is
\(
H^*(BF_4; \Z_3) \cong
\Z_3[x_{36}, x_{48}] \otimes
\left( \Z_3[x_4 , x_8] \otimes \left\{ 1, x_{20}, x_{20}^2 \right\}
+ \Lambda(x_9) \otimes \Z_3[x_{26}] \otimes\left\{1, x_{20}, x_{21}, x_{25} \right\}
\right).
\)
The generators can be chosen to be related by the Steenrod power operations
at $p=3$, $P^i : H^n(BF_4 ; \Z_3) \to H^{n+4i} (BF_4; \Z_3)$, as
\(
\begin{array}{lllll}
x_8=P^1 x_4 && x_9 =\beta x_8= \beta P^1 x_4 && x_{20}=P^3 P^1 x_4 \\
x_{21}=\beta P^3 P^1 x_4 && x_{25}=P^4 \beta P^1 x_4 && x_{26}= \beta P^4\beta P^1 x_4
\end{array}
\)
and $x_{48}=P^3 x_{36}$. 
If we restrict to degrees $\leq 11$ then we have the truncated polynomial
\(
H^*(BF_4; \Z_3) \cong  \Z_3[x_4, x_8] + \Lambda(x_9).
\label{trun1}
\)

\vspace{3mm}
The classes coming from $B{\rm Spin}(9)$ are just the Stiefel-Whitney classes
in the $\Z_2$ case and the Pontrjagin classes (reduced mod 3) in the integral
($\Z_3$ case). These are actually not much different from the classes
of $B{\rm Spin}(11)$. Explicitly, at $p=2$ 
the cohomology ring of $B{\rm Spin}(9)$ is given by the polynomial
ring \cite{extra} 
\( 
H^*(B{\rm Spin}(9); \Z_2)=\Z_2 \left[ w_4, w_6, w_7,
w_8, w_{16}'\right], 
\label{bspin92}
\) 
where $w_i$ is the restriction of the
universal Stiefel-Whitney class, and $w_{16}'$ is the
Stiefel-Whitney class $\w_{16}(\Delta_{{\rm Spin}(9)})$ of the spin
representation $\Delta_{{\rm Spin}(9)}: {\rm Spin}(9) \to O(16)$.
At $p=3$, $H^*(B{\rm Spin}(9);\Z_3)$ is generated by the first
four Pontrjagin classes \cite{Toda}
\(
H^*(B{\rm Spin}(9);\Z_3)=\Z_3[p_1, p_2, p_3, p_4],~~~{\rm deg}(p_i)=4i \; .
\label{bspin93}
\)
Let us look at $\Z_3$ coefficients. From (\ref{trun1}) and (\ref{bspin93}) we see that 
$H^9(B{\rm Spin}(9);\Z_3) =0$ while $H^9(BF_4;\Z_3) \neq 0$, which implies that 
the map $H^9(BF_4;\Z_3) \to H^9(B{\rm Spin}(9);\Z_3)$ cannot be injective. 
Therefore, at $p=3$ the Serre spectral sequence is not trivial. In the case of 
$\Z_2$, the situation is reversed, this time in degree eight: 
$H^8(B{\rm Spin}(9);\Z_2)\neq 0$ and $H^8(BF_4;\Z_2)=0$.

\vspace{3mm}
Now we proceed with the uniqueness by applying the results in \cite{KST}. 
The cohomology of ${\mathbb O} P^2$ is $H^*({\mathbb O} P^2; C)=C[x]/x^3$, 
$|x|={\rm deg}\hspace{0.5mm}x=8$,
as an algebra. Then, requiring that the Serre fibering ${\mathbb O} P^2 \to M^{27} \to Y^{11}$
be trivial over $C$ implies for the $E_2$-term 
\(
E_2 =H^*(Y^{11};C) \otimes_C C[x]/x^3.
\)
Now the $E_9$ term is $E_{|x|+1}=E_2$ and the fibering is nontrivial
if and only if we have a nonzero differential
$
d_9(1 \otimes x) \neq 0 
$. 
If $d_9 ( 1 \otimes x) = a \otimes 1 \neq 0$ then
$
0= d_9 ( 1 \otimes x^3)= 3 ( a \otimes x^2)
$.
Hence the characteristic of $C$ must not be relatively prime to 3, the
degree of the ideal in the cohomology ring of ${\mathbb O} P^2$. Therefore, we have

\begin{prop}
The Serre spectral sequence for the fiber bundle ${\mathbb O} P^2 \to M^{27} \to Y^{11}$
is nontrivial only for cohomology with $\Z_3$ coefficients. 
\end{prop}

\noindent We will make use of this and also say more in section \ref{top terms} -- see 
theorem \ref{d9thm} and the discussion around it.

\begin{prop}
If $Y^{11}$ admits a {\it String} structure then so does $M^{27}$ provided that
there is no contribution from the degree four class from $BF_4$. 
\label{cond}
\end{prop}

\proof
We have the ${\mathbb O} P^2$ bundle over $Y^{11}$ with total space $M^{27}$
\( 
\xymatrix{M^{27} \ar[r]^{\tilde{f}} \ar[d]_{\pi} & B{\rm Spin}(9)
\ar[d]^{Bi}\\ Y^{11} \ar[r]_{f} & BF_4}, 
\) 
which gives the
decomposition $TM^{27}=\pi^*TY^{11}\oplus {\tilde f}^*T$, and so the
tangential Pontrjagin class is 
\( 
p_1(M^{27})= \pi^*\left(
p_1(Y^{11})+f^*p_1(T) \right). 
\) 
In the case $Y^{11}$ is a
3-connected $BO\langle 8 \rangle$-manifold, we have that
$H^4(Y^{11};\Z)$ is free and $\pi^* : H^4(Y^{11};\Z) \to
H^4(M^{27};\Z)$ is an isomorphism.
Thus $M^{27}$ is also a $BO\langle 8 \rangle$-manifold if and only
if $f^*\overline{x}_4=0 \in H^4(Y^{11};\Z)$, where $\overline{x}_4
\in H^4(BF_4;\Z)$ is the generator. Therefore we have shown that $M^{27}$
is {\it String} if and only if $G_4$ in M-theory gets no contribution from
$BF_4$.
\endproof

\noindent {\bf Remarks}

\noindent {\bf 1.} The quantization condition for the field strength $G_4$ in
M-theory is known \cite{Flux}. Since this field does not seem to get a contribution 
from a class in $BF_4$, the condition in Proposition \ref{cond} seems
reasonable. In some sense we could view the presence of such a degree 
four class as an anomaly which we have just cured. 
%Alternatively, 
%one can discover that this is not as serious as it might seem--- see
%the more complete discussion in section \ref{effect on PF}.

\noindent {\bf 2.} We connect the above discussion back to cobordism
groups. While there is no transfer map from $\Omega_{11}^{\langle 8
\rangle}(BF_4)$ to $\Omega_{27}^{\langle 8 \rangle}$, there is a
transfer map after killing $\overline{x}_4$ \cite{Klaus}. Denoting
by \footnote{This is the analog of the {\it String} group when $G={\rm
Spin}$, in the sense that it is the 3-connected cover.} $BF_4\langle
\overline{x}_4\rangle$ the corresponding classifying space that
fibers over $BF_4$, killing $\overline{x}_4$ is done by pulling back
the path fibration $PK(\Z,4) \to K(\Z, 4)$ with a map $\overline{x}_4
: BF_4 \to K(\Z, 4)$ realizing $\overline{x}_4$. The corresponding
transfer map is $\Omega_{11}^{\langle 8 \rangle}(BF_4\langle
\overline{x}_4 \rangle) \to \Omega_{27}^{\langle 8 \rangle}$.

\vspace{3mm}
Next, for the higher structures we have

\begin{prop}
\noindent {\bf 1.} In order for $M^{27}$ to admit a Fivebrane structure, 
the second Pontrjagin class of $Y^{11}$ should be the negative of that
of ${\mathbb O} P^2$, i.e. $p_2(TY^{11})= -p_2(T {\mathbb O} P^2)=- 6u$. 

\noindent {\bf 2.} $\widehat{A} (M^{27})=0$, irrespective of whether 
or not the $\widehat{A}$-genus of $Y^{11}$ is zero.

\noindent {\bf 3.} $\Phi_W(M^{27})=0$.

\noindent {\bf 4.} $\Phi_{\rm ell}(M^{27})=0$.
\label{ellm27}
\end{prop}
\proof
For part (1) note that if $Y^{11}$ admits a Fivebrane structure 
then $M^{27}$ does not necessarily admit such a structure.
This is because the obstruction to having a Fivebrane structure is
$\frac{1}{6}p_2$ \cite{SSS2} but 
we know that $\frac{1}{6}p_2({\mathbb O} P^2)=u \neq 0$. 
However, we can choose $Y^{11}$ appropriately so that it 
conspires with ${\mathbb O} P^2$ to cancel the obstruction and lead
to a Fivebrane structure for $M^{27}$. 
Noting that the tangent bundles are related as $TM^{27}= TY^{11} \oplus T{\mathbb O} P^2$,
then considering the degree eight
part of the formula (see \cite{Mil})
$p(E \oplus F)= \sum p(E) p(F)$ mod $2-$torsion,
we get for our spaces
\(
p_2(TY^{11} \oplus T {\mathbb O} P^2)= p_1(TY^{11}) p_1(T {\mathbb O} P^2)
+ p_2(TY^{11}) + p_2(T {\mathbb O} P^2)~~~{\rm mod~}2{\rm -torsion}.
\)
Since we have $p_1(T {\mathbb O} P^2)=0$, then requiring that
$p_2(TM^{27})=0$ leads to the constraint that
$p_2(TY^{11})+ p_2(T {\mathbb O} P^2)=0$ modulo 2-torsion.

For part (2) we use the multiplicative property of the $\widehat{A}$-genus
for Spin fiber bundles to get
\(
\widehat{A}(M^{27})= \widehat{A} (Y^{11}) 
\widehat{A} ({\mathbb O} P^2).
\)
Since the $\widehat{A}$-genus of ${\mathbb O} P^2$ is zero then the result follows.

For part (3) we use a result of Ochanine \cite{Och1}. Taking the total space 
$M^{27}$ and the base $Y^{11}$ to be closed oriented manifolds, and since 
the fiber ${\mathbb O} P^2$ is a Spin manifold and the structure group $F_4$ 
of the bundle is compact, then the multiplicative property of the genus can be 
applied
\(
\Phi_W (M^{27})= \Phi_W ({\mathbb O} P^2) \Phi_W (Y^{11}).
\)
We proved in Theorem \ref{genus} that $\Phi_W ({\mathbb O} P^2)=0$, so it 
follows immediately that $\Phi (M^{27})$ is zero regardless of whether
or not $\Phi_W (Y^{11})$ vanishes. Even more, $\Phi_W(Y^{11})$ is zero
because $Y^{11}$ is odd-dimensional.
\footnote{ However, see the case when
$Y^{11}$ is a circle bundle at the end of this section.}

For part (4) we use the fact that the fiber is Spin and the structure group
$F_4$ is compact and connected so we can apply the multiplicative 
property of the elliptic genus \cite{Och1}
\(
\Phi_{\rm ell}(M^{27})= \Phi_{\rm ell}(Y^{11}) \Phi_{\rm ell}({\mathbb O} P^2).
\)
In this case the genus for the fiber is not zero (see Proposition \ref{ellop2})
but the elliptic genus of $Y^{11}$ is zero, again because of dimension.
Therefore $\Phi_{\rm ell}(M^{27})=0$. 
\endproof

\vspace{3mm}
We next consider the relation between the Ochanine genera of 
the base and of the total space. 

\vspace{3mm}
Having the Ochanine genera for $S^1$ and $X^{10}$, we now proceed
to determine the corresponding genus for the eleven-dimensional manifold $Y^{11}$.

\begin{prop}
Let $Y^{11}$ be an eleven-dimensional Spin manifold which is the total space
of a circle bundle over a ten-dimensional Spin manifold $X^{10}$. Then the Ochanine
genus of $Y^{11}$ is 
\(
\Phi_{\rm och} (Y^{11})= \Phi_{\rm och} (X^{10})\cdot \alpha (S^1)\;. 
\)
\label{ochy11}
\end{prop} 
\vspace{-.5cm}
\proof
\vspace{-.5cm}
Unlike other genera, the Ochanine genus does not in 
general enjoy a multiplicative property on fiber bundles. However, 
in the special case when the fiber is the circle with a $U(1)$ action 
$\Phi_{\rm och}$ does become multiplicative on the circle bundle 
\cite{KS}. We simply apply the result for $S^1 \to Y^{11} \to X^{10}$
to get 
\(
\Phi_{\rm och} (Y^{11}) = \Phi_{\rm och} (X^{10}) \cdot \Phi_{\rm och} (S^1).
\) 
With $\Phi_{\rm och}(S^1)= \alpha (S^1)$ the degree one generator in $KO_*({\rm pt})$, 
the result follows.
\endproof

Now that we have the Ochanine genus for $Y^{11}$, we go back and consider 
the original questions of finding the Ochanine genus of $M^{27}$, given that
of $Y^{11}$. 

\begin{theorem}
The Ochanine genus of the total space $M^{27}$ of an ${\mathbb O} P^2$ bundle 
over an eleven-dimensional compact Spin manifold $Y^{11}$, which is a
circle bundle over a ten-dimensional Spin manifold $X^{10}$, is 
\(
\Phi_{\rm och} (M^{27})=\Phi_{\rm och} ({\mathbb O} P^2) \cdot \Phi_{\rm och} (X^{10}) \cdot 
\alpha (S^1)\;,
\)
where $\Phi_{\rm och} ({\mathbb O} P^2)$ is given in Theorem \ref{ochop2} and 
$\Phi_{\rm och} (X^{10})$ is given as follows:
If $k(X^{10})=0 \in \Z_2$
then $\Phi_{\rm och}(X^{10})=\alpha (X^{10})$, while if 
$k(X^{10})=1 \in \Z_2$ then in $KO_{10}\otimes \Z_2$ we have 
\(
\Phi_{\rm och}(X^{10})= \alpha (X^{10}) + \eta^2 \mu (q + q^9 + q^{25} + \cdots)\;.
\)
\label{theorem och}
\end{theorem}

\proof
\vspace{-0.5cm}
As mentioned in the proof of Proposition \ref{ochy11} above, $\Phi_{\rm och}$ 
is not in general multiplicative for fiber bundles. Again, interestingly, we are in 
a special case where such a  property holds \cite{KS}. It is so 
because the dimension of the fiber ${\mathbb O} P^2$ is a multiple of 4, the structure
group $F_4$ is a compact connected Lie group,  and the base $Y^{11}$ 
is a closed Spin manifold. Applying to the fiber bundle 
${\mathbb O} P^2 \to M^{27} \to Y^{11}$, and using proposition \ref{ochy11},
 then gives the formula in the 
theorem.
\endproof

\noindent {\bf Remark.} The circle in Theorem \ref{theorem och} is the one with
the nontrivial/nonbounding/supersymmetric/Ramond-Ramond
 Spin structure. 
% as in Lemma 
%\ref{ochs1}
%(cf. the remark after that Lemma).

%%%%%%%%%%%%%%
\subsubsection{Topological terms in the lifted action}
%%%%%%%%%%%%%%
\label{top terms}

Having motivated and then constructed ${\mathbb O} P^2$ bundles in M-theory, we now
turn to the discussion of some of the consequences. The most obvious question
from a physics point of view is to characterize the corresponding `theory' in
27 dimensions. We will not be able to achieve that, but we will be able to 
characterize some of the terms in the would-be action up in 27 dimensions.
In the absence of a clear handle, we take the most economical 
approach and concentrate on the topological terms, which in any case are the
terms we can trust. We 
also make some remarks on other terms as well.

%\subsection{Topological terms}

%\label{top terms}

\vspace{3mm}
The simplest topological term coming from ${\mathbb O} P^2$
at the rational level would be some differential form of degree 
sixteen. This could also be decomposable, i.e. a wedge product
of differential forms of lower degrees such that the total degree 
is 16. We should seek forms that naturally occur on ${\mathbb O} P^2$.
Looking at the question from a 27-dimensional perspective,
a Kaluza-Klein mechanism comes to mind. We do not attempt 
to discuss this problem fully here but merely provide some
possibilities that are compatible with the structures that
we have. In dimensional reduction from ten and eleven
dimensions to lower dimensions, holonomy plays an important
role as it gives some handle on the differential forms involved,
as well as on supersymmetry.

\vspace{3mm}
From the cohomology of ${\mathbb O} P^2$, the possible topological terms
generated from this internal space come from $X_i \in H^i({\mathbb O} P^2)$
for $i=8, 16$, so that their linear combination generates a candidate
degree sixteen term 
\( 
\rho_{16}:= a X_{16} + b X_8^2,
\label{r16}
\) 
where $X_8$ and $X_{16}$ are eight- and sixteen-forms, respectively, and 
$a$ and $b$ are some parameters. 

\vspace{3mm}
\noindent {\bf Remarks}

\noindent {\bf 1.} Since the degree 16 generator is built out of the degree
8 generator, namely the first is proportional to $u^2$ and the second is $u$,
then equation (\ref{r16}) is redundant as $X_{16}$ is really built out of
$X_{8}^2$. Thus equation (\ref{r16}) should be replaced by
$\rho_{16}=b X_8^2$.

\noindent {\bf 2.} In terms of the generator $u$ of $H^8({\mathbb O} P^2;\Z)$, the 
expression at the integral level should be
\(
\rho_{16} = \alpha u^2,
\)
with $\alpha \in \mathbb{Q}$.

\noindent {\bf 3.} The term $\rho_{16}$ would be thought of as a degree 16 analog of 
the one loop term $I_8$ in M-theory and type IIA string theory from \cite{DLM}.
It would appear as a topological term 
in the action, rationally as 
\(
S_{(27)}^{\rm top}= \int_{M^{27}} L_{(27)}^{\rm top} =
\int_{M^{27}} \rho_{16} \wedge L_{(11)^{\rm top}}\;,
\)
where $L_{(11)}^{\rm top}$ is the topological Lagrangian in 
eleven dimensions given by
\(
L_{(11)}^{\rm top}= \frac{1}{6}G_4 \wedge G_4 \wedge C_3 - I_8 \wedge C_3\,.
\)
Then we have 
\(
S_{(27)}^{\rm top}
= \int_{Y^{11}} L_{(11)}^{\rm top}
\int_{{\mathbb O} P^2} \rho_{16} 
\nonumber\\
= \alpha \int_{Y^{11}} L_{(11)}^{\rm top}
\nonumber\\
= \alpha \hspace{0.5mm} S_{(11)}^{\rm top}\; .
\label{integrate}
\)

\noindent {\bf 1.} At the rational level we can thus use $\omega_8$ to build 
a ${\rm Spin}(9)$-invariant degree sixteen expression  
$
\rho_{16}^{\R}= \omega_8 \wedge \omega_8
$
that we integrate and insert as part of the action as $\int_{{\mathbb O} P^2} \rho_{16}^{\R}$. 
%
%\noindent {\bf 2.} Assume that there are 
%fields $\mathcal{F}_8$ and $\mathcal{F}_{16}$ in the 27-dimensional 
%`theory' with potentials ${\mathcal C}_7$ and ${\mathcal C}_{15}$. 
% In the dimensional reduction on ${\mathbb O} P^2$ to eleven dimensions,
%a natural ${\rm Spin}(9)$-invariant 
%ansatz for the fields may be taken, at the rational level, to be 
%\(
%{\mathcal F}_8 = \omega_8, ~~~~~~~~~~~~~~~~~~ 
%{\mathcal F}_{16} =\omega_8 \wedge \omega_8\;,
%\label{cal F}
%\)
%and similar expressions at the integral level in terms of ${\cal J}_8$. Note that
%since $\omega_{16}$ is essentially the volume form, then such an ansatz is 
%the analog of the Freund-Rubin ansatz \cite{FR} in the 
%reduction of eleven-dimensional supergravity to lower dimensions. 

%

%

\vspace{3mm}
The integration of $\rho_{16}$ over ${\mathbb O} P^2$ in the second step of equation
(\ref{integrate}) requires the existence of a fundamental class $[{\mathbb O} P^2]$ for
the Cayley plane. The Cayley 8-form ${\cal J}_8$ allows for such an 
evaluation at the rational and integral level. The next question is about
torsion. The existence of such a fundamental class at that level 
is neither automatic nor obvious. In order to state the following result we recall some notation.
Let $\beta: H^i(Y^{11}; \Z_3) \to H^{i+1}(Y^{11}; \Z)$ be the Bockstein homomorphism
corresponding to the reduction modulo 3, $r_3: \Z \to \Z_3$,
i.e. associated to the short exact sequence 
$
0 \to \Z_3 \to \Z_9 \to \Z_3 \to 0
$ 
and $P_3^1: H^j(Y^{11}; \Z_3) \to H^{j+4}(Y^{11}; \Z_3)$ be the Steenrod 
reduced power operation at $p=3$. Then we have

\begin{theorem}
A fundamental class exists provided that $\beta P_3^1 x_4=0$, where
$x_4$ is the mod 3 class on $Y^{11}$ pulled back from $BF_4$ via the 
 classifying map.
 \label{d9thm}
\end{theorem}

\proof
Consider the fiber bundle $E \to Y^{11}$ with fiber ${\mathbb O} P^2$ and
structure group $F_4$. There is a universal bundle of this type.
 ${\mathbb O} P^2$ bundles over $Y^{11}$ are 
pullbacks of the universal bundle 
\( 
{\mathbb O} P^2=F_4/{\rm Spin}(9) \longrightarrow
B{\rm Spin}(9) \longrightarrow BF_4 
\label{main} 
\) 
by the classifying map $f :
Y^{11} \to BF_4$.
Since $BF_4$ is path-connected and ${\mathbb O} P^2$ is connected then we
can apply the Serre spectral sequence to the fibration (\ref{main}). 
We consider two cases for the coefficients of the cohomology: 
$\Z_p$(or any field in general), $p$ a prime, and $\Z$ coefficients.

\vspace{3mm}
\noindent {\it Coefficients in $\Z_p$:} The important primes are $p=2, 3$
as these are the torsion primes of $F_4$. 
For $p=2$ the inclusion map $i:{\rm Spin}(9) \hookrightarrow
F_4$ induces a map on the classifying spaces so that $H^*(B {\rm
Spin}(9); \Z_p)$ is a free $H^*(BF_4;\Z_p)$-module on generators 
$1, x, x^2$ with $x \in H^8 (B{\rm Spin}(9); \Z_p)$ the 
universal Leray-Hirsch generator that
maps to $x \in H^8 ({\mathbb O} P^2;\Z_p)$. 
Here we use the fact \cite{Clear} that 
the Serre spectral sequence for a fiber bundle $F \to E \to B$
collapses if and only if the corresponding Poincar\'e series 
${\mathcal P}(-):=\sum_{n \geq 0} t^n {\rm dim}_{\Z_p} H^n(-; \Z_p)$
satisfies
${\mathcal P}(E)={\mathcal P}(F){\mathcal P}(B)$.
In our case the Serre spectral sequence of
(\ref{main}) collapses \cite{Klaus}. This follows from the equality
of the corresponding Poincar\'e polynomials 
\bea 
\frac{\cP (B{\rm Spin}(9))}{\cP (BF_4)}&=&
\frac{(1-t^4)^{-1}(1-t^6)^{-1}(1-t^7)^{-1}(1-t^8)^{-1}(1-t^{16})^{-1}}
{(1-t^4)^{-1}(1-t^6)^{-1}(1-t^7)^{-1}(1-t^{16})^{-1}(1-t^{24})^{-1}}
\nonumber\\
&=&\frac{1-t^{24}}{1-t^8}=1+t^8+t^{16}, 
\eea 
which is just the
Poincar\'e polynomial $\cP ({\mathbb O} P^2)$ of the Cayley plane. This implies
that the Leray-Hirsch theorem holds, i.e. that the map $H^*({\mathbb O} P^2)
\otimes H^*(BF_4) \to H^*(B{\rm Spin}(9))$ is an isomorphism of
$H^*(BF_4)$-modules. This implies in particular that $H^*(B{\rm
Spin}(9))$ is a free $BF_4$-module on $1, x, x^2$, where $x$ is
either $w_8$ or $w_8 + w_4^2$. The Wu formula with $w_1=w_2=0$ for
both cases gives that $Sq^1x=Sq^2 x= Sq^3 x=Sq^5x=0$ 
so that 
\( Sq \hspace{0.5mm}x= x + Sq^4 x + Sq^6 x+ Sq^7 x+ x^2. 
\) The elements $x_4$, $Sq^2
x_4$, $Sq^3 x_4 \in H^*BF_4$ are mapped to the elements $w_4,
w_6=Sq^2w_4, w_7=Sq^3w_4 \in H^*B{\rm Spin}(9)$. The Leray-Hirsch
theorem holds for the universal bundle, and consequently for all 
${\mathbb O} P^2$ bundles \cite{Klaus}.

For $p=3$ the argument is similar except that now the generators in 
degrees 4 and 8 are related as $p_1=\overline{p}_1$ and 
$p_2={\overline p}_2 + {\overline p}_1^2$, respectively  
\cite{Toda}. Here $p_i$ are the Pontrjagin classes (see the appendix).

\vspace{3mm}
\noindent {\it Coefficients in $\Z$:} We would like to find the differentials for 
\(
H^*(B{\rm Spin}(9);\Z) \Longleftarrow
H^* \left( BF_4, H^*({\mathbb O} P^2; \Z) \right).
\)
The class $u$ maps under the differential to a $\Z_3$ class of degree 9 
which we will call $\alpha$. The lowest degree class on the fiber is 
$x_8$, so the differentials begin with $d_9$. The differential is $d_9$ on $x_8$ so that
the class is $\beta P_3^1 x_4$, where $x_4$ is the mod 3 class on
$Y^{11}$ coming from $BF_4$
\(
Y^{11} \longrightarrow BF_4 \longrightarrow K(\Z_3, 9).
\)
We thus have a 3-torsion class 
of ${\mathbb O} P^2$ bundles. The obstruction in $H^9(Y^{11}; \Z)$ coming 
from $H^9(BF_4; \Z)$ is zero if and only if there exists a degree 16
class, say $\rho_{16}$, that restricts on each fiber to the fundamental
class. 
\endproof

Thus the vanishing of $d_9$ provides us with a fundamental
class which we use to integrate over ${\mathbb O} P^2$.

\endproof

\begin{remark}
The Pontrjagin classes $p_2$ and $p_4$ of ${\mathbb O} P^2$ are divisible by three.
There is always a class in $M^{27}$ that restricts on the fiber to three times the
generator of the cohomology of ${\mathbb O} P^2$. 
\end{remark}

\section{Connection to Cobordism and Elliptic Homology} 
%%%%%%%%%%%%%%%%%%%%%%
\label{top}

%%%%%%%%%%%%%%%%%%
\subsection{Cobordism and boundary theories}
%%%%%%%%%%%%%%%%%%%%%
\label{cobor}

In this section we consider the question of extension of the theories in
 eleven and twenty-seven dimensions to bounding theories in twelve
 and twenty-eight dimensions, respectively, assuming the spaces to be {\it String} 
 and taking into account the $F_4$ bundles. 
As mentioned in the introduction,
our discussion will make contact with a version of
elliptic cohomology constructed by Kreck and Stolz \cite{KS}. In
that paper the emphasis was on the Spin case corresponding
geometrically to quaternionic projective plane $\H P^2$ bundles, but
the authors assert the existence of a $BO\langle 8 \rangle$ version
corresponding to octonionic projective plane ${\mathbb O} P^2$ bundles. Let
us denote this theory by $E^{\langle 8 \rangle}$ or, equivalently,
by $E^{{\mathbb O}}$.

\vspace{3mm}
%Note that we can also consider the String condition, 
%discussed above in section \ref{bund},
%from an
%eleven-dimensional point of view. 

We consider the {\it String} condition from an eleven-dimensional 
point of view.
One point that we utilize
is that $\Omega_{11}^{\rm spin}({\rm pt})$, the Spin cobordism group
in eleven dimensions, is zero. This means that any
eleven-dimensional Spin manifold bounds a twelve-dimensional one. It
is also the case that the $BO \langle 8 \rangle$ cobordism group
$\Omega_{11}^{\langle 8 \rangle}({\rm pt})$ is zero \cite{Giam1}, so that the
extension from an eleven-dimensional {\it String} manifold to the
corresponding boundary is unobstructed. Thus, if the space $Y^{11}$
in which M-theory is defined admits a {\it String} structure
then this always bounds a
twelve-dimensional {\it String} manifold $Z^{12}$.

\vspace{3mm} Generalized cohomology theories can, in fact, be 
obtained as
quotients of cobordism (see \cite{KS1} for some exposition on this
for physicists)
by classic results \cite{CF}. For instance, Spin cobordism
$\Omega_*^{\rm spin}=\Omega_*^{\langle 4 \rangle}$ 
is closely related to real K-theory $KO$, a fact
we used in section \ref{bund}.
For a space $X$, $KO^*({\rm pt})$ can be made into an $\Omega^{\rm
spin}_*$-module and there is an isomorphism of $KO^*(X)$ 
with $\Omega_*^{\rm spin}(X)
\otimes_{\Omega_*^{\rm spin}} KO^*({\rm pt})$. As we have seen, this is 
related to the mod 2
index of the Dirac operator with values in real bundles in ten
dimensions which appears in the mod 2 part of the partition function
\cite{DMW}. There is an analogous construction for elliptic cohomology,
where there the starting point is $\Omega_*^{\langle 8 \rangle}$.
This fact is related to the
elliptic refinement of the mod 2 index which then has values in a
real version of elliptic cohomology \cite{KS1}.

%%%%%%%%%
\subsection{Cobordism of $BO\langle 8 \rangle$-manifolds with fiber ${\mathbb O} P^2$}
%%%%%%%%%
Now we go back to our main discussion of relating the 
cobordisms of the eleven- and twenty-seven-dimensional 
theories together with the $F_4$-${\mathbb O} P^2$ bundles. Thus we 
 are led to the study of the cobordism  groups $\Omega_i^{\langle 8 \rangle}(BF_4)$
 for $i=11$ and 27. We will also be interested in relating these two groups.

\vspace{3mm}
We have an 11-dimensional base manifold $Y^{11}$, assumed to admit a 
${\rm String}(11)$ structure, with an ${\mathbb O} P^2$
bundle such that the total space is $M^{27}$ and the structure group is $F_4$.
Let ${\mathcal I} \in \Omega_{27}^{\langle 8 \rangle}$ be the ideal generated by 
elements of the form $[M^{27}] -[{\mathbb O} P^2] [Y^{11}]$ where, as before,
$M^{27} \to Y^{11}$ is a fibration with fiber ${\mathbb O} P^2$ and structure 
group $F_4$. We have 

\begin{prop}
Let $Y^{11}$ be a compact manifold with a {\it String} structure on which M-theory
is taken, and let $M^{27}$ be the {\it String} manifold on which the 27-dimensional 
theory is taken,
realizing the Euler triplets geometrically. Then such 27-manifolds $M^{27}$ are
in the ideal $\mathcal I$ of $\Omega_{27}^{\langle 8 \rangle}$ generated by
${\mathbb O} P^2$ bundles.
\end{prop}
Our setting is given in the following diagram
\(
  \xymatrix{
 {\mathbb O} P^2
  \ar[rr]
  &&
   M^{27}
   \ar[rrdd]^{f'}
   \ar[dd]_{\pi}
   &&
  &&
   \\
   \\
   &&
   Y^{11}
    \ar[rr]_{f}
  &&
N \;.
  }
  \label{diag2}
  \)
First we ignore the structure group and consider $N$ to be a point. As in Section
\ref{cobor}, let $\Omega_*^{\langle 8 \rangle}$ be the cobordism ring of manifolds
with $w_1=w_2=\frac{1}{2}p_1=0$. This ring has only 2-torsion and 3-torsion, with 
the 3-torsion being a $\Z_3$ summand in dimensions 3, 10, and 13 (this is known 
only up to roughly dimension 16).

\vspace{3mm}
Note that cobordim groups $\Omega_*^{\langle n \rangle}$ arise 
as homotopy groups of the Thom spectra $MO \langle n \rangle$, 
in the sense that the former groups are the homotopy groups of the
spectra (this is general for any type of cobordism).  
Hence the Thom spectrum for
the {\it String} cobordism ring is $MO \langle 8 \rangle$, and
$\Omega_*^{\langle 8 \rangle}=\pi_* \left( MO \langle 8 \rangle \right)$.
We can actually gain information about $\Omega_*^{\langle 8 \rangle}$ 
by looking at topological modular forms. This is due to the following 
fact.
Let $MO \langle 8 \rangle \to tmf$ be any multiplicative map
whose underlying genus is the Witten genus. Then the induced map on
the homotopy groups $\pi_* MO \langle 8 \rangle \to \pi_* tmf$ is surjective
\cite{Hop}. The low-dimensional homotopy groups of $tmf$ are \cite{Hop}
\footnote{Here we prefer to use the notation for cyclic groups used in 
homotopy theory, e.g. $\Z/2$ in place of $\Z_2$. We hope this will be clear.}

\bigskip
\begin{center}
\begin{tabular}{|c|c|c|c|c|c|c|c|c|c|c|c|c|c|c|c|c|c|c|c|}\hline
$k$ & $0$ & $1$ & $2$ & $3$ & $4$ & $5$ & $6$ & $7$ & $8$ 
&$9$ & $10$ & $11$ & $12$ & $13$ & $14$ & $15$  \\ \hline
$\pi_{k}\tmf$ & $\Z$ & $\Z/2$ & $\Z/2$  & $\Z/24$ & $0$ & $0$ &
$\Z/2$ & 0 & $\Z \oplus \Z/2$& $(\Z/2)^{2}$ & $\Z/6$ & $0$ & $\Z$ & $\Z/3$ & $\Z/2$ &
$\Z/2$ 
\\ \hline
\end{tabular}
\end{center}

\bigskip

\noindent The 2-primary components ${}_{(2)}\Omega_*^{\langle 8 \rangle}$ of 
$\Omega_*^{\langle 8 \rangle}$ are given by \cite{Giam1} (see also
\cite{Klaus} \cite{Thom})

\bigskip
\begin{center}
\begin{tabular}{|c|c|c|c|c|c|c|c|c|c|c|c|c|c|c|c|c|c|c|c|cl}\hline
$k$ & $0$ & $1$ & $2$ & $3$ & $4$ & $5$ & $6$ & $7$ & $8$   
& $9$ & $10$ & $11$ &
$12$ & $13$ & $14$ & $15$  & $16$ 
\\ \hline
${}_{(2)}\Omega_k^{\langle 8 \rangle}$ & $\Z$ & $\Z/2$ & $\Z/2$  & $\Z/8$ & $0$ & $0$ &
$\Z/2$ & 0 & $\Z \oplus \Z/2$ &
$(\Z/2)^{2}$ & $\Z/2$ & $0$ & $\Z$ & $0$ & $\Z/2$ &
$\Z/2$ & $(\Z)^2$
\\ \hline
\end{tabular}
\end{center}
 
\bigskip

\noindent By comparing the two tables, we can indeed see the `missing' $\Z/3$ factors.

\vspace{3mm}
Note that in dimension 11, the result of 
\cite{Giam1} implies that $\Omega_{11}^{\langle 8 \rangle}=0$ since the 2-primary 
part is zero and there is no torsion in that dimension. There does not seem to
be a computation for dimensions as high as 27. This implies that the map
\(
\varrho : \Omega_{11}^{\langle 8 \rangle}({\rm pt}) \longrightarrow 
\Omega_{27}^{\langle 8 \rangle} ({\rm pt})
\label{zero}
\)
is a map whose domain is 0, and is thus not interesting.

\vspace{3mm}
We next allow the structure group $F_4$ so that there is a map from
$Y^{11}$ to its classifying space $BF_4$. Thus we are considering 
$N=BF_4$ and the classifying map to be $f$ in (\ref{diag2}). In this case,
instead of the map $\varrho$ we will consider the map
\bea
\varrho' : \Omega_{11}^{\langle 8 \rangle}(BF_4) &\longrightarrow &
\Omega_{27}^{\langle 8 \rangle} (BF_4)
\\
{[ Y^{11}, f]} &\longmapsto & {[M^{27}, f']}\;,
\eea
which maps bordism classes of 11-manifolds, together with a map 
$f$ to $BF_4$, to bordism classes of 27-manifolds together with 
a map $f'$ to $BF_4$. Now both the
domain and the range are in general non-empty unless certain condition are
applied. 

\vspace{3mm}
\noindent {\bf Remarks}

\noindent {\bf 1.} The classifying space $BF_4$ has at least interesting
degree four cohomology. However, we have seen 
that for the {\it String} condition to be multiplicative
on ${\mathbb O} P^2$ bundles then we must kill $x_4$ coming from $BF_4$. This would
then mean that we should in this case consider $BF\langle x_4 \rangle$ instead
of $BF_4$.

\noindent {\bf 2.} Killing $x_4$ as above would lead to the rational 
homotopy type
\(
BF_4 \langle x_4 \rangle \sim S^{12} \times {\rm higher~spheres}\;, 
\)
so that the first homotopy is in dimension 12. This then would mean
that should consider 
$\Omega_{11}^{\langle 8 \rangle} \left(BF_4 \langle x_4 \rangle \right)$,
which is zero, by dimension.

\noindent {\bf 3.} If we use $BF_4 \langle x_4 \rangle$ instead of $BF_4$, 
then this might cause some problems for the description of the fields of
M-theory in terms of ${\mathbb O} P^2$ bundles, since there we used the Lie
group $F_4$ on the nose. In other words, unlike the case for compact 
$E_8$ in eleven dimensions, $F_4$ appears not merely 
topologically, but via representation theory. However,  
compare to the arguments in \cite{Sgerbe} for the $E_8$ model of the
$C$-field in M-theory.
It should be checked that the representations coming from the Lie 2-group 
$F_4 \langle x_4 \rangle$ respect the discussion in section \ref{count}.

\vspace{3mm}
We can actually say more about the extensions of the $F_4$ bundle.
We have

\begin{prop}
The $F_4$ bundle on a {\it String} manifold $Y^{11}$ can be extended to 
$Z^{12}$ where $\partial Z^{12}=Y^{11}$.
\label{hi}
\end{prop}
\proof
We look for cobordism obstructions.
 Extending the bundle would be obstructed by 
$\Omega_{11}^{\langle n \rangle} (BF_4 )$. Since the homotopy type of
$F_4$ is $(3, 11, 15, 23)$ then that of $BF_4$ is $(4, 12, 16, 24)$ so that
up to dimension 11 the classifying space $BF_4$ has the homotopy type
of $K(Z, 4)$, much the same as $E_8$ does (and in fact all exceptional 
Lie groups except $E_6$) in that range. Now we reduce the problem
to checking whether 
$\Omega_{11}^{\langle n \rangle} \langle( K(\Z, 4)\rangle)$ is zero. This
is indeed so by calculations of Stong \cite{Stong}, for $n=4$, and Hill (\cite{Hill},
motivated by this question), 
for $n=8$. 
\endproof

\vspace{3mm}
Let $T_{27}^{\langle 8 \rangle}(BF_4)$ be the subgroup of  
$\Omega_{27}^{\langle 8 \rangle}(BF_4)$ consisting of bordism classes
$[M^{27}, f \circ \pi]$, i.e. the classes that factor through the base $Y^{11}$.
It could happen that some of the classes $[Y^{11}, f]$ of the bordism group
of the base are zero.
Let $\widetilde{T}_{27}^{\langle 8 \rangle}(BF_4)$  be the subgroup
whose elements satisfy the additional assumption that $[Y^{11}, f]=0$
in $\Omega_{11}^{\langle 8 \rangle}(BF_4)$.
Corresponding to the diagram (\ref{diag}) there is a classifying map 
 \( 
 \psi : \Omega_{11}^{\langle 8
\rangle} (BF_4) \longrightarrow \Omega_{27}^{\langle 8 \rangle} ({\rm pt})
\label{psi}
\)
which takes the class $[Y^{11}, f]$ to the class $[M^{27}=f^*E]$.
The image $T_{27}^{\langle 8 \rangle}={\rm im}\hspace{0.5mm} \psi$ of this map 
is the set of total
spaces of ${\mathbb O} P^2$ bundles in $\Omega_{27}^{\langle 8 \rangle}$.   
If we forget the
classifying map $f$ then instead of (\ref{psi}) we can map
 \( 
 \lambda: \Omega_{11}^{\langle 8 \rangle}
(BF_4) \longrightarrow \Omega_{11}^{\langle 8 \rangle} ({\rm pt})\;,
\label{pi}
\) 
where now
the class $[Y^{11}, f]$ lands in the class $[Y^{11}]$ by simply
forgetting $f$. Obviously, the kernel of $\lambda$ makes up the classes
$[Y^{11}, f]$ which map to $[Y^{11}]$ that are zero in
$\Omega_{11}^{\langle 8 \rangle}$. Such classes $[Y^{11}, f]$ map
under $\psi$ to total spaces of ${\mathbb O} P^2$ bundles with zero-bordant
bases in $\Omega_{11}^{\langle 8 \rangle}$. It is clear that
$\psi ({\rm ker}\hspace{0.5mm} \lambda)$ 
is the subgroup $\widetilde{T}_{27}^{\langle 8 \rangle}$. That is, we have
\bea
T_{27}^{\langle 8 \rangle}&:=& {\rm im}\hspace{0.5mm} \psi = \left\{ 
{\rm total~spaces~of~} {\mathbb O} P^2{\rm ~bundles~ in~}
\Omega_{27}^{\langle 8 \rangle}({\rm pt})  \right\}
\\
{\widetilde T}_{27}^{\langle 8 \rangle}&:=& \psi (\ker \lambda)
\label{t}
\nonumber\\
&=& 
\left\{ {\rm total~spaces~of~}{\mathbb O} P^2{\rm ~bundles~
with~zero~bordant~base~
in~}\Omega_{27}^{\langle 8 \rangle}({\rm pt})  \right\}.
\label{tt}
\eea

\vspace{3mm}
Note that, as mentioned above, the 2-primary 
part of $\Omega_{n}^{\langle 8 \rangle}$ for 
$n \leq 16$ is calculated in \cite{Giam1}. For $n=11$ this is zero.
This implies that the kernel of $\lambda$ is all of 
$ \Omega_{11}^{\langle 8 \rangle} (BF_4)$, i.e. all cobordism classes
of total spaces have zero bordant bases. Then we have
\begin{prop}
$T_{27}^{\langle 8 \rangle}$ and ${\widetilde T}_{27}^{\langle 8 \rangle}$
coincide for base {\it String} manifolds of dimension eleven. 
\end{prop}
There are two cases to consider in order to determine whether or not 
the above spaces are trivial:

\noindent {\bf 1.} If $\Omega_{27}^{\langle 8 \rangle}$ turns out to be zero, then
the map $\psi$ will be trivial in that degree.

\noindent {\bf 2.} If it turns out that $\Omega_{27}^{\langle 8 \rangle}\neq 0$, 
then the map $\psi$ is not trivial. It would then mean that 
 $T_{27}^{\langle 8 \rangle}
 ={\widetilde T}_{27}^{\langle 8 \rangle}\neq \emptyset$. 
 However, looking carefully at the map $\psi$ we notice that its domain
 is zero. This is because the homotopy type of $F_4$ is $K(\Z,3)$ up
 to dimension ten, so that the homotopy type of $BF_4$ is $K(\Z,4)$ up
 to dimension eleven. This means that 
 $\Omega_{11}^{\langle 8 \rangle}(BF_4)=\Omega_{11}^{\langle 8 \rangle}
 ((K\Z, 4))=0$. This then implies that the map $\psi$ is trivial. 
 In modding out by the corresponding equivalence
to form 
\(
E_{27}^{{\mathbb O}}=E_{27}^{\langle 8 \rangle}= \Omega_{27}^{\langle 8 \rangle}/
T_{27}^{\langle 8 \rangle}\;,
\label{div}
\)
we simply get 

\begin{prop}
The homology theory is just the bordism ring 
$E_{27}^{{\mathbb O}}=\Omega_{27}^{\langle 8 \rangle}$. 
\label{equality}
\end{prop}
%We will come back to this below.

\vspace{3mm}
\noindent { \bf Remarks}

\noindent {\bf 1.} Proposition \ref{equality} implies that in dimension 27 we do not get
anything smaller or simpler than bordism.  

\noindent {\bf 2.} The two spaces (\ref{t}) and (\ref{tt}) have been characterized in
the quaternionic case, i.e. when the fiber is $\H P^2$ with structure group $PSp(3)$, as
\bea
T_{27}^{\langle 4 \rangle} &=& \ker (\alpha)
\\
{\widetilde T}_{27}^{\langle 4 \rangle} &=& \ker (\Phi_{\rm och}),
\eea
i.e. as the kernels of the Atiyah invariant in \cite{Stolz} and the Ochanine 
genus in \cite{KS}, respectively.  We see that in our case, 
$\alpha(M^{27})=0$, but $\Phi_{\rm och}(M^{27})$ is not necessarily 
zero. This provides another justification for the calculations leading to
theorem \ref{theorem och}. In fact, we can use the nontriviality of the 
Ochanine genus to check whether or not the homology theory is empty.
Since, using Theorem \ref{theorem och}, 
we can find a 27-dimensional manifold $M^{27}$ with
$\Phi_{\rm och}(M^{27})\neq 0$, the Spin cobordism 
group is nonzero $\Omega_{27}^{\langle 4 \rangle}\neq0$. Consequently,
we have the following result for the corresponding {\it String} cobordism group. 

\begin{theorem}
$\Omega_{27}^{\langle 8 \rangle} \neq 0$\;.
\end{theorem}

\paragraph{Remark.} Alternatively, the theorem can proved using 
information about $tmf$. Since the orientation 
map from $M{\rm String}=MO\langle 8 \rangle$ to $tmf$ is
surjective \cite{AHS} then  it is enough to know that the 
homotopy group of $tmf$ in dimension 27 is nonzero. 
Indeed,  
\footnote{I thank Mike Hill for pointing out the $\Z/3$ summand in this homotopy group.}
 at least
 $\pi_{27}(tmf) \supset \Z/3$, so that 
$\Omega_{27}=\pi_{27}(M{\rm String}) \neq 0$.

%\noindent {\bf 3.} 
%The problem for us will be to characterize the spaces $T$ 
%for the octonionic case. 
\vspace{3mm}
In \cite{Thom}, the Witten genus was proposed as
a candidate for the replacement of $\alpha$ in the octonionic case, so that
\(
T_{27}^{\langle 8 \rangle} ({\rm pt}) = \ker (\alpha^{{\mathbb O}}) :=\ker (\Phi_W). 
\)
Indeed, we have shown in Proposition \ref{ellm27} that the Witten genus 
is zero for our 27-dimensional manifolds, which are ${\mathbb O} P^2$ bundles. 
The extension of the the `new Atiyah invariant' $\alpha^{{\mathbb O}}$ would be to
a `new Ochanine genus' 
\(
\Phi_{\rm och}^{{\mathbb O}} : \Omega_*^{\langle 8 \rangle} \longrightarrow \Q[E_4, E_6][[q]], 
\)
i.e. to the power series ring over rationalized coefficients of level 1 elliptic
cohomology, such that the constant term is the Witten genus. We have seen
in theorem \ref{genus} that the Witten genus of ${\mathbb O} P^2$ is zero, so that
in the current context, the constant term is zero. We do not know what 
the higher terms are, and so they can conceivably be nonzero. 
The `new Ochanine genus' is expected to be related to $K3$-cohomology. Such 
a theory has not yet been explicitly constructed but it should exist.

\vspace{3mm}
Define the functor 
$X \to \Omega_*^{\langle 8 \rangle}(X)/{\mathcal I}$, where ${\cal I}$ is the ideal 
introduced in the beginning of this section. The question is 
whether this is a generalized (co)homology theory. 
The desired homology theory
$E_n^{{\mathbb O}}$ is formed by dividing $\Omega_*^{\langle 8 \rangle}$ by
$\widetilde{T}_n$ and inverting the primes 2 and 3 \cite{Thom}.
However, there is one extra condition required, which is the
invertibility of the element $v={\mathbb O} P^2$. By taking the limit in
 \(
E_n^{{\mathbb O}}(X)[{\mathbb O} P^2]^{-1}= {\lim}_j E_{n+16j}^{{\mathbb O}}(X) 
\) 
over the
sequence of homomorphisms given by multiplying by ${\mathbb O} P^2$ the
resulting theory is 
\( 
e\ell \ell_*^{{\mathbb O}} (X)=E_*^{{\mathbb O}}(X)[{\mathbb O} P^2]^{-1}=\bigoplus_{k \geq 0}
\Omega_{*+ 16k}(X)/\sim, 
\) 
where the equivalence relation $\sim$ is
generated by identifying $[Y,f]\in \Omega_*^{\langle 8 \rangle}(X)$
with $[M, f\circ \pi] \in \Omega_{*+16k}^{\langle 8 \rangle}(X)$ for an
${\mathbb O} P^2$ bundle $\pi : M \to Y$, with structure group 
${\rm Isom}\hspace{0.5mm}{\mathbb O}P^2=F_4$, i.e. the total 
space of an ${\mathbb O} P^2$ bundle is identified with
its base. A full construction of this theory is not yet achieved by homotopy
theorists but it is believed that this should be possible in principle. 
We mentioned towards the end of Section \ref{cobor} 
that $KO^*({\rm pt})$ can be made into an $\Omega_*^{\rm spin}$-module
and the existence of an isomorphism relating $KO^*(X)$ and $KO^*({\rm pt})$.
The octonionic version of Kreck-Stolz theory is arrived
at by replacing $KO^*(\rm pt)$ by $e\ell \ell_n^{{\mathbb O}}(\rm pt)$, i.e. 
\(
\Omega_*^{\langle 8 \rangle} (X) \otimes_{\Omega_*^{\langle 8
\rangle}} e \ell \ell_*^{{\mathbb O}}(\rm pt) \longrightarrow e\ell \ell_*^{{\mathbb O}}(X) 
\) 
is an isomorphism away from the
primes 2 and 3 \cite{Thom}. 

\vspace{3mm}
\noindent {\bf Remarks.}

\noindent {\bf 1.} The model for elliptic homology in fact involves indefinitely
higher cobordism groups in increments of 16, 
\(
e \ell \ell_{11}^{{\mathbb O}}(Y^{11})=\bigoplus_{k \geq 0} \Omega_{11+ 16k}/\sim, 
\label{high}
\)
where $\sim$ is an equivalence that provides a correlation between
topology in M-theory and topology in dimensions $27, 43, \cdots,
11+16k, \cdots, \infty$. We have two points to make:

\begin{itemize}
\item The first bundle with total space an ${\mathbb O} P^2$ bundle over
$Y^{11}$ is related to Ramond's Euler multiplet.

\item As the pattern continues in higher and higher dimensions, one
is tempted to seek physical interpretations for such theories as
well. While this direction is tantalizing, we do not pursue it in
this paper.
\end{itemize}

\noindent {\bf 2.} There is another homology theory that one can form, namely 
by identifying the image of $\psi$ with the trivial bundle as in \cite{KS}. The 
construction is analogous. The advantage here is that we do not kill $\O P^2$,
as dividing by $T$ has the effect of killing the fiber. 

\vspace{3mm}
We have seen connections between eleven-dimensional M-theory and the 
putative theory in twenty-seven dimensions.
If the latter theory in twenty-seven dimensions is fundamental, then it should 
ultimately be studied also without restricting to the relation to M-theory.
This is analogous to the case of M-theory itself in relation to ten-dimensional
type IIA string theory. Since M-theory is, as far as we know, a fundamental 
theory, then it should be (and it is being) studied without necessarily
assuming a circle bundle for the eleven-dimensional manifold.
In other words, what about 27-dimensional manifolds that are not
the total space of ${\mathbb O} P^2$ bundles over eleven-manifolds? Hence

\paragraph{Proposal.}  
To study the bosonic theory as a fundamental theory in twenty-seven
dimensions we should also consider modding out by the equivalence 
relation (the ideal).

\vspace{3mm}
\noindent For example, extension problems can be studied in this way.

%%%%%%%%%
\subsection{Families}
%%%%%%%%%
It is desirable to consider the ${\mathbb O} P^2$ bundle as a family problem 
of objects on the fiber of $M^{27}$ parametrized by points 
in the base $Y^{11}$. The family of these 16-dimensional {\it String} 
manifolds will define an element of the cobordism group 
\(
{MO \langle 8 \rangle}^{-16} (Y^{11}). 
\)

\paragraph{Remarks}

\noindent {\bf 1.} We have seen in section \ref{conseq} that the total space of an ${\mathbb O} P^2$
bundle is not necessarily {\it String} even if $Y^{11}$ is {\it String}. 
However, we do get a family of {\it String} manifolds provided we kill the
degree four class pulled back from $BF_4$ (see Prop. \ref{cond}).

 \noindent {\bf 2.} Unfortunately, genera are multiplicative on fiber bundles so that the vanishing 
of $\Phi_W({\mathbb O} P^2)$ will force the Witten genus of $M^{27}$ to be zero as
well. Also taking higher and higher bundles -- so as to get fibers of dimensions higher
than 16-- as in (\ref{high}) will not help in
making the Witten genus nonzero. $tmf$ is the home of the parametrized
version of the Witten genus, but we do not see modular forms in this
picture.  This is to be contrasted with the $\H P^2$ case where the Witten genus
is  $E_4/288$. 

\noindent {\bf 3.} Nevertheless, the elliptic genus $\Phi_{\rm ell}$
of ${\mathbb O} P^2$ is not zero, so the total space will not automatically have 
a zero elliptic genus. However, elliptic genera are defined for Spin 
manifolds of dimension divisible by 4. Our base space $Y^{11}$ is 
eleven-dimensional and so will automatically have zero elliptic
genus. This also applies for the Witten genus. One way out of this
is instead to consider the bounding twelve-dimensional theory,
i.e. the extension of the topological terms from $Y^{11}=\partial Z^{12}$ 
to $Z^{12}$ as in \cite{Flux}. If we also take a 28-dimensional
coboundary for $M^{27}$, i.e. $\partial W^{28}=M^{27}$, we would then 
have  
 \(
  \xymatrix{
  {\mathbb O} P^2
  \ar[rr]
  &&
   M^{27}~
    \ar@{^{(}->}[rr]
   \ar[dd]_{\pi}
   &&
  W^{28}
  \ar[dd]_{\pi}
  \ar[rrdd]
  &&
   \\
   \\
   &&
   Y^{11}~
    \ar@{^{(}->}[rr]
  &&
  Z^{12}
      \ar[rr]_{f}
      &&
BF_4 \; .
  }
\)
Such an extension would involve cobordism obstructions. The manifolds
extend nicely, as $\Omega_{11}^{\langle n \rangle}=0$ for both $n=4$ (Spin)
and $n=8$ ({\it String}). The bundles also extend as shown in Proposition
\ref{hi}. It is tempting to propose that the theories should be defined 
on the $(12+16m)$-dimensional spaces, and then restriction to the 
boundaries would be a special instance.

\vspace{3mm}
We have provided evidence for some relations between M-theory and an 
octonionic version of Kreck-Stolz elliptic
homology. Strictly speaking, both theories are
conjectural, and we hope that this contribution motivates more
active research both on completing the mathematical construction of
this elliptic homology theory (part of which is outlined in
\cite{Thom}) as well as making more use of the connection to
M-theory. In doing so, we even hope that M-theory itself would in
turn give more insights into the homotopy theory.

\vspace{3mm}
In closing we hope that further investigation will help shed more 
light on the mysterious appearance of the exceptional groups
$E_8$ and $F_4$ and to give a better understanding of their
role in M-theory.

\vspace{1cm}

%%%%%%%%%%%%%%%%%%%%%%%%%%%%%%%%%%%%%%%%%%%%%%%%%%
{\bf \large Acknowledgements}
%%%%%%%%%%%%%%%%%%%%%%%%%%%%%%%%%%%%%%%%%%%%%%%%%%

\vspace{2mm}
\noindent 
The author thanks the American Institute of Mathematics for hospitality and the 
`` Algebraic Topology and Physics" SQuaRE program participants Matthew Ando, 
John Francis, Nora Ganter, Mike Hill, Mikhail Kapranov, Jack Morava, Nitu Kitchloo, 
and Corbett Redden for very useful discussions. 
The author would like to thank the Hausdorff Institute for Mathematics in Bonn
for hospitality and the organizers of the ``Geometry and Physics" Trimester Program
at HIM for the inspiring atmosphere during the writing of this paper.  Special 
thanks are due to Matthew Ando, Mark Hovey, Nitu Kitchloo, Matthias Kreck, 
and Pierre Ramond for helpful remarks and encouragement, to 
Stephan Stolz for useful comments on the draft, and to Arthur Greenspoon for  
many useful editorial suggestions.


\begin{thebibliography}{999}
%%%%%%%%%%%%%%%%%%%%%%%%%%%%%%

%\bibitem{AM}
%K. Abe and M. Matsubara,
%{\it Invariant forms on the exceptional symmetric spaces 
%$FII$ and $EIII$},
%Transformation group theory (Taej\u on, 1996), 3--16, Korea Adv. 
%Inst. Sci. Tech., Taej\u on, 1996. 

\bibitem{Adams}
J. F. Adams,
Lectures on exceptional Lie groups,
University of Chicago Press, Chicago, IL, 1996.

\bibitem{AE}
A. Adams and J. Evslin,
{\it The loop group of $E_8$ and K-theory from 11d},
J. High Energy Phys. {\bf 0302} (2003) 029,
[{\tt arXiv:hep-th/0203218}].

%\bibitem{AG}
%L. Alvarez-Gaum\'e and P. H. Ginsparg, 
%{\it The Structure of Gauge and Gravitational Anomalies}, 
%Annals Phys. {\bf 161} (1985) 423.

\bibitem{ABP2}
D. W. Anderson, E. H. Brown, Jr., and F. P. Peterson, 
{\it ${\rm SU}$-cobordism, ${\rm KO}$-characteristic numbers, and the Kervaire invariant},
 Ann. of Math. {\bf(2) 83} (1966) 54--67. 

\bibitem{AHS}
 M. Ando, M. J. Hopkins, and N. P. Strickland, 
 {\it Elliptic spectra, the Witten genus and the theorem of the cube},
  Invent. Math. {\bf 146} (2001), no. 3, 595--687. 

%\bibitem{As}
%H. Aslaksen,
%{\it Restricted homogeneous coordinates for the Cayley projective plane}, 
%Geom. Dedicata {\bf 40} (1991), no. 2, 245--250.

%
%\bibitem{At}
%M. F. Atiyah, 
%{\it Riemann surfaces and spin structures},
%Ann. Sci. \'Ecole Norm. Sup. (4) {\bf 4} (1971) 47--62. 


 \bibitem{ABS}
M. F. Atiyah, R. Bott, and A. Shapiro,
{\it Clifford modules},
Topology {\bf 3} (1964) suppl. 1, 3--38.
%
%\bibitem{AH}
%M. F. Atiyah and F. Hirzebruch, 
%{\it Vector bundles and homogeneous spaces},
%Proc. Sympos. Pure Math., Vol. III pp. 7--38, AMS , Providence, R.I., 1961.
%
%\bibitem{AS}
%M. F. Atiyah and I. M. Singer, 
%{\it The index of elliptic operators. V}, 
%Ann. of Math. {\bf (2) 93} (1971) 139--149. 

\bibitem{Baez}
J. Baez,
{\it The octonions},
Bull. Amer. Math. Soc. {\bf 39} (2002) 145-205. Erratum ibid. {\bf 42} (2005) 213,
[{\tt arXiv:math/0105155}] [math.RA].
%
%\bibitem{Bar}
%C.~B{\"a}r, {\it Real {K}illing spinors and holonomy},
%Commun. Math. Phys. {\bf 154} (1993) 509--521.
%
%\bibitem{Bar2}
%C.~B{\"a}r, {\it Elliptic symbols}, Math. Nachr. {\bf 201} (1999), 7--35. 

\bibitem{BM1}
D. Belov and G. M. Moore, {\it Holographic action for the self-dual
field},
 [{\tt arXiv:hep-th/0605038}].

\bibitem{BM2}
D. Belov and G. M. Moore,{\it Type II actions from 11-dimensional
Chern-Simons theories}, \newline
[{\tt arXiv:hep-th/0611020}].
%
%\bibitem{Ben}
%M. Bendersky,
%{\it Applications of the Ochanine genus}, 
%Math. Z. {\bf 206} (1991), no. 3, 443--455. 
%
%\bibitem{Besse}
%A. L. Besse, Einstein manifolds, Springer-Verlag, Berlin, 1987.

\bibitem{Borel}
A. Borel, 
{\it Sur l'homologie et la cohomologie des groupes de Lie compacts connexes},
Amer. J. Math. {\bf 76} (1954)  273--342. 

%\bibitem{BH}
%A. Borel and F. Hirzebruch,
%{\it Characteristic classes and homogeneous spaces. I},
%Amer. J. Math. {\bf 80} (1958) 458--538. 

%\bibitem{BP}
%C. Brada and F. Pecaut-Tison, 
%{\it G\'eom\'etrie du plan projectif des octaves de Cayley},
%Geom. Dedicata {\bf 23} (1987) no. 2, 131--154. 

%\bibitem{Brown}
%E. H. Brown, Jr.,
%{\it Generalizations of the Kervaire invariant}, 
%Ann. of Math. {\bf (2) 95} (1972) 368--383. 

%\bibitem{BG}
%R. B. Brown and A. Gray, 
%{\it Riemannian manifolds with holonomy group $Spin(9)$},
%Differential geometry (in honor of Kentaro Yano), eds. S. Kobayashi et al., 
%pp. 41--59. Kinokuniya, Tokyo, 1972. 

%\bibitem{CENT}
%A. Casher, F. Englert, H. Nicolai, and A. Taormina,
%{\it Consistent superstrings as solutions of the $D=26$ bosonic 
%string theory},
%Phys. Lett. {\bf B162} (1985) 121.

\bibitem{CF}
P. Conner and E. Floyd, {\it The  relation of cobordism to $K$-theories}, 
Lecture Notes in Mathematics, no. {\bf 28} Springer-Verlag, Berlin-New York 1966.

\bibitem{CJS}
E. Cremmer, B. Julia, and J. Scherk,
{\it Supergravity theory in eleven-dimensions},
Phys. Lett. {\bf B76} (1978) 409-412.

\bibitem{DFM}
E. Diaconescu, D. S. Freed, and G. Moore, {\it The M-theory 3-form
and $E_8$ gauge theory}, Elliptic cohomology, 44--88, 
London Math. Soc. Lecture Note Ser., 342, Cambridge Univ. Press, Cambridge, 2007, 
[{\tt arXiv:hep-th/0312069}].

\bibitem{DMW}
E.~Diaconescu, G.~Moore, and E.~Witten, {\it $E_8$ gauge theory, and
a derivation of K-Theory from M-Theory}, Adv. Theor. Math. Phys.
{\bf 6} (2003) 1031, [{\tt arXiv:hep-th/0005090}].

%\bibitem{Duff1}
%M. J. Duff, 
%{\it $E_8 \times SO(16)$ symmetry of $d = 11$ supergravity?}, 
%in Quantum Field Theory and Quantum Statistics,  
%eds. I. A. Batalin, C. Isham and G. A. Vilkovisky, Adam Hilger, Bristol,  UK 
%1986.

\bibitem{Duff2}
M. J. Duff,
{\it M-theory (the theory formerly known as strings)},
Int. J. Mod. Phys. {\bf A11} (1996) 5623-5642,
[{\tt 	arXiv:hep-th/9608117v3}].

\bibitem{DLM}
M. J. Duff, J. T. Liu, and R. Minasian, {\it Eleven dimensional
origin of string/string duality: A one loop test}, Nucl. Phys. {\bf
B452} (1995) 261, [{\tt arXiv:hep-th/9506126}].

%\bibitem{E8T}
%J. Evslin,
%{\it From $E_8$ to $F$ via $T$},
%J. High Energy Phys. {\bf 0408} (2004) 021,
%[{\tt arXiv:hep-th/0311235}].

%
%\bibitem{JFO}
%J. Figueroa-O'Farrill,
%{\it  A geometric construction of the exceptional Lie algebras $F_4$ and $E_8$},
%\newline
%[{\tt arXiv:0706.2829v}] [math.DG] .

%\bibitem{FR}
%P. G. O. Freund and M. A. Rubin,
%{\it  Dynamics of dimensional reduction},
% Phys. Lett. {\bf B97} (1980) 233-235.


\bibitem{Fried}
T. Friedrich, {\it Weak $Spin(9)$-structures on 16-dimensional
Riemannian manifolds}, 
Asian J. Math. {\bf 5} (2001) 129--160, 
[{\tt arXiv:math/9912112}] [math.DG].

\bibitem{Giam1}
 V. Giambalvo, 
{\it On $\langle 8\rangle$-cobordism},
Illinois J. Math. {\bf 15} (1971) 533--541; erratum ibid 
{\bf 16} (1972) 704.

%\bibitem{GG}
%A. Gray and P. Green, 
%{\it Sphere transitive structures and the triality automorphism},
%Pac. J. Math. {\bf 34} (1970), 83-96.
 

\bibitem{GSW}
M. B. Green, J. H. Schwarz, and E. Witten, 
Superstring theory Vol. 2, second edition,
Cambridge University Press, Cambridge, 1988.

%\bibitem{GKRS}
%B. Gross, B. Kostant, P. Ramond, and S. Sternberg, 
%{\it The Weyl character formula, the half-spin representations, and equal rank subgroups}
%Proc. Natl. Acad. Sci. USA {\bf 95} (1998), no. 15, 8441--8442,
%[{\tt arXiv:math/9808133}] [math.RT].

\bibitem{Har}
F. R. Harvey,
Spinors and Calibrations, Academic Press, Boston, 1990.


%\bibitem{HSV}
%R. Held, I. Stavrov, and B. Van Koten,
%{\it (Semi-)Riemannian geometry of (para-)octonionic projective planes},
%[{\tt arXiv:math/0702631}] [math.DG]. 
 

\bibitem{Hill}
M. Hill, 
{\it The String bordism of $BE_8$ and $BE_8\times BE_8$ through dimension 14},
[{\tt arXiv:0807.2095}] [math.AT].

%
%\bibitem{Inertia}
%F. Hirzebruch,
%{\it On Steenrod's reduced powers, the index of inertia, and the Todd genus}, 
%Proc. Nat. Acad. Sci. U.S.A. {\bf 39} (1953) 951--956. 

%\bibitem{Hir}
%F. Hirzebruch,
%Topological methods in algebraic geometry,
%Springer-Verlag, 1978. 

\bibitem{HiS}
F. Hirzebruch and P. Slodowy, {\it Elliptic 
genera, involutions, and homogeneous spin manifolds}, 
Geom. Dedicata  {\bf 35}  (1990),  no. 1-3, 309--343.

\bibitem{Hit}
N. Hitchin, 
{\it Harmonic spinors}, 
Advances in Math. {\bf 14} (1974) 1--55. 

\bibitem{Hop}
M. J. Hopkins,
{\it Algebraic topology and modular forms},
Proceedings of the ICM, Beijing 2002, vol. {\bf 1}, 283--309,
[{\tt arXiv:math/0212397v1}] [math.AT].

\bibitem{HS}
G. Horowitz and L. Susskind, {\it Bosonic M-theory}, J. Math. Phys.
{\bf 42}(2001) 3152, \newline 
[{\tt arXiv:hep-th/0012037}].

\bibitem{Hov1}
M. Hovey, {\it Spin bordism and elliptic homology},
Math. Z. {\bf 219} (1995), no. 2, 163--170.

%\bibitem{Hull}
%C. M. Hull,
%{\it Duality and the signature of space-time},
%J. High Energy Phys. {\bf 9811} (1998) 017,
%[{\tt arXiv:hep-th/9807127}].

%\bibitem{Hus}
%D. Husemoller, 
%Fibre bundles, Third edition, Springer-Verlag, New York, 1994.

%\bibitem{IPW}
%C. J. Isham, C. N. Pope, and N. P. Warner,
%{\it Nowhere-vanishing spinors and triality rotations in $8$-manifolds}, 
%Class. Quant. Gravity {\bf 5} (1988), no. 10, 1297--1311. 

%\bibitem{Julia1}
%B. Julia,
%{\it Group disintegrations}, in Superspace and supergravity,
%S.W. Hawking and M. Ro{\v c}ek (eds.), Cambridge Univ. Press,
%Cambridge , UK, 1981.

%\bibitem{Keur}
%A. Keurentjes,
%{\it   The group theory of oxidation},
%Nucl. Phys. {\bf B658} (2003) 303-347,\newline
%[{\tt arXiv:hep-th/0210178}].

\bibitem{Klaus}
S. Klaus, {\it Brown-Kervaire invariants},
PhD thesis, University of Mainz, Shaker Verlag, Aachen, 1995.

\bibitem{Klaus2}
S. Klaus, 
{\it The Ochanine $k$-invariant is a Brown-Kervaire invariant}, 
Topology {\bf 36} (1997), no. 1, 257--270. 
%
%\bibitem{KNS}
%K. Koepsell, H. Nicolai, and H. Samtleben,
%{\it An exceptional geometry for d=11 supergravity?},
%Class. Quant. Grav. {\bf 17} (2000) 3689-3702,
%[{\tt arXiv:hep-th/0006034}].

\bibitem{KST}
A. Kono, H. Shiga, and M. Tezuka, 
{\it A note on the cohomology of a fiber space whose 
fiber is a homogeneous space},
Quart. J. Math. Oxford Ser. (2) {\bf 40} (1989), no. 159, 291--299.


%\bibitem{Kos}
%B. Kostant, 
%{\it A cubic Dirac operator and the emergence of Euler number 
%multiplets of representations for equal rank subgroups},
% Duke Math. J. {\bf 100} (1999), no. 3, 447--501.
 
\bibitem{KS}
M. Kreck and S. Stolz, 
{\it ${\bf H}{\rm P}^2$-bundles and elliptic homology}, 
Acta Math. {\bf 171} (1993) 231--261. 

\bibitem{KS1}
I.~Kriz and H.~Sati, {\it M Theory, type IIA superstrings, and
elliptic cohomology}, Adv. Theor. Math. Phys. {\bf 8} (2004) 345,
[{\tt arXiv:hep-th/0404013}].

\bibitem{KS2}
I.~Kriz and H.~Sati, {\it Type IIB string theory, S-duality and
generalized cohomology}, Nucl. Phys. {\bf B715} (2005) 639, [{\tt
arXiv:hep-th/0410293}].

\bibitem{KS3}
I.~Kriz and H.~Sati, {\it Type II string theory and modularity},
J. High Energy Phys. {\bf 08} (2005) 038, [{\tt arXiv:hep-th/0501060}].

%\bibitem{LW}
%N. D. Lambert and P. C. West,
%{\it Coset symmetries in dimensionally reduced bosonic string theory},
%Nucl. Phys. {\bf B615} (2001) 117-132,
%[{\tt arXiv:hep-th/0107209}].

%\bibitem{Land1}
%G. Landweber,
%{\it Harmonic spinors on homogeneous spaces},
%Represent. Theory {\bf 4} (2000) 466-473,
%[{\tt arXiv:math/0005056v1}] [math.DG].

\bibitem{Li}
A. Lichnerowicz,
{\it Spineurs harmoniques},
C. R. Acad. Sci. Paris {\bf 257} (1963) 7--9. 

\bibitem{24}
M. Mahowald and M. Hopkins, 
{\it The structure of 24-dimensional manifolds having normal bundles which lift to $B{\rm O}[8]$},
in Recent progress in homotopy theory (Baltimore, MD, 2000), 
89--110, Contemp. Math., 293, Amer. Math. Soc., Providence, RI, 2002. 

\bibitem{TDMW}
V. Mathai and H. Sati,
{\it Some relations between twisted K-theory and $E_8$ gauge theory},
J. High Energy Phys.  {\bf 0403} (2004) 016,
[{\tt arXiv:hep-th/0312033}].


\bibitem{Clear}
J. McCleary, User's Guide to Spectral Sequences, 
Publish or Perish, Wilmington, Delaware, 1984.

%\bibitem{Mich}
% P. W. Michor, 
%  Gauge theory for fiber bundles,
% Bibliopolis, Napoli, 1991.
  


\bibitem{Mil} J. Milnor and J. Stasheff, Characteristic
classes, Princeton University Press, Princeton, NJ, 1974.
%
%\bibitem{Nic1}
%H. Nicolai,
%{\it $D = 11$ supergravity with local ${\rm SO}(16)$ invariance},
% Phys. Lett. {\bf B187} (1987) 316.

\bibitem{Och0}
S. Ochanine,
{\it Signature modulo $16$, invariants de Kervaire g\'en\'eralis\'es 
et nombres caract\'eristiques dans la $K$-th\'eorie r\'eelle},
MŽm. Soc. Math. France (N.S.) {\bf 5} 1980/81. 


\bibitem{Och1}
S. Ochanine, {\it Genres elliptiques \'equivariants}, in
Elliptic curves and modular forms in algebraic topology (Princeton, NJ, 1986), 107--122, 
Lecture Notes in Math., 1326, Springer, Berlin, 1988. 

\bibitem{Och2}
S. Ochanine, 
{\it Elliptic genera, modular forms over $K{\rm O}\sb *$ and the 
Brown-Kervaire invariant},
Math. Z. {\bf 206} (1991), no. 2, 277--291. 

\bibitem{Ram1}
T. Pengpan and P. Ramond,
{\it M(ysterious) Patterns in ${\rm SO}(9)$},
Phys. Rept. {\bf 315} (1999) 137-152,
[{\tt arXiv:hep-th/9808190}].

\bibitem{extra}
D. Quillen, 
{\it The ${\rm mod}$ $2$ cohomology rings of extra-special $2$-groups and the spinor groups},
 Math. Ann. {\bf 194} (1971) 197--212.

\bibitem{Ram2}
P. Ramond,
{\it Boson-fermion confusion: The string path to supersymmetry},
Nucl. Phys. Proc. Suppl. {\bf 101} (2001) 45-53,
[{\tt arXiv:hep-th/0102012}].

\bibitem{Ram3}
P. Ramond, {\it Algebraic dreams}, Meeting on Strings and Gravity: Tying the Forces Together, Brussels, Belgium, 19-21 Oct 2001, [{\tt arXiv:hep-th/0112261}].

\bibitem{Rey}
S.-J.Rey, {\it Heterotic M(atrix) strings and their interactions},
Nucl. Phys. {\bf B502} (1997) 170, \newline
[{\tt arXiv:hep-th/9704158}].

\bibitem{Compact}
H. Salzmann, D. Betten, T. Grundh{\"o}fer, 
H. H{\"a}hl, R. L{\"o}wen, 
and M. Stroppel, 
Compact Projective Planes,  
Walter de Gruyter \& Co., Berlin, 1995.  

\bibitem{S1}
H. Sati, {\it M-theory and Characteristic Classes}, J. High Energy
Phys. {\bf 0508} (2005) 020, \newline
[{\tt arXiv:hep-th/0501245}].

\bibitem{S2}
H. Sati, {\it Flux quantization and the M-theoretic characters},
Nucl. Phys. {\bf B727} (2005) 461, \newline
[{\tt arXiv:hep-th/0507106}].

\bibitem{S3}
H. Sati, {\it Duality symmetry and the form-fields in M-theory}, J.
High Energy Phys. {\bf 0606} (2006) 062, \newline
[{\tt arXiv:hep-th/0509046}].


\bibitem{S4}
H. Sati, {\it The Elliptic curves in string theory, gauge theory,
and cohomology}, J. High Energy Phys. {\bf 0603} (2006) 096, [{\tt
arXiv:hep-th/0511087}]

\bibitem{Sgerbe}
H. Sati, {\it $E_8$ gauge theory and gerbes in string theory},
[{\tt arXiv:hep-th/0608190}].

\bibitem{Stwist}
H. Sati, {\it On higher twist in string theory},
[{\tt arXiv:hep-th/0701232}].

\bibitem{Sloop}
H. Sati, {\it Loop group of $E_8$ and targets for spacetime}, 
[{\tt arXiv:hep-th/070123}].

\bibitem{KSpin}
H. Sati,
{\it An approach to anomalies in M-theory via $K{\rm Spin}$},
J. Geom. Phys. {\bf 58} (2008) 387, \newline
[{\tt arXiv:0705.3484}] [hep-th]. 

\bibitem{SSS1} H. Sati, U. Schreiber and J. Stasheff,
{\it $L_\infty$-connections and applications to String- and
Chern-Simons $n$-transport}, in {\it Recent Developments in QFT},
eds. B. Fauser et al., Birkh{\"a}user, Basel (2008), [{\tt
arXiv:0801.3480}] [math.DG].

\bibitem{SSS2}
H. Sati, U. Schreiber, and J. Stasheff,
{\it Fivebrane structures}, to appear in Rev. Math. Phys., 
[{\tt arXiv:0805.0564}] [math.AT].
 
 \bibitem{Stolz}
 S. Stolz, 
{\it Simply connected manifolds of positive scalar curvature},
Ann. of Math. {\bf (2) 136} (1992), no. 3, 511--540. 
 
% \bibitem{Conj}
% S. Stolz,
%{\it  A conjecture concerning positive Ricci curvature and the Witten genus},
%Math. Ann. {\bf 304} (1996), no. 4, 785--800. 
 
%\bibitem{ST}
%S. Stolz and P. Teichner,
%{\it What is an elliptic object?} in Topology, geometry and quantum field theory, 
%London Math. Soc. LNS 308, Cambridge Univ. Press 2004, 247-343.

\bibitem{StongNotes}
R. E. Stong, 
Notes on cobordism theory, Princeton University Press, Princeton, N.J., 1968.

\bibitem{Stong}
R. Stong, {\it Calculation of $\Omega _{11}^{spin}(K(\Z,4))$},
in Unified String Theories,  M. Green and D. Gross, eds.,  pp. 430--437,
World Scientific 1986.

\bibitem{Thom}
C. Thomas, {\it Elliptic Cohomology}, Kluwer Academic/Plenum Publishers, New York, 1999.

\bibitem{Toda}
H. Toda, {\it Cohomology ${\rm mod}$ $3$ of the classifying space $BF\sb{4}$ of the 
exceptional group $F\sb{4}$},  J. Math. Kyoto Univ. {\bf 13} (1973) 97--115. 

\bibitem{Town}
P. K. Townsend,
{\it Four lectures on M-theory}, Summer School in High Energy Physics and 
Cosmology Proceedings, E. Gava et. al (eds.),
 Singapore, World Scientific, 1997,
[{\tt arXiv:hep-th/9612121v3}].


\bibitem{Dynamics}
E. Witten,
{\it String theory dynamics in various dimensions},
Nucl. Phys. {\bf B443} (1995) 85-126, \newline
[{\tt arXiv:hep-th/9503124v2}].


\bibitem{Flux}
E. Witten, {\it On Flux quantization in M-theory and the effective
action}, J. Geom. Phys. {\bf 22} (1997) 1, [{\tt arXiv:hep-th/9609122}].

\bibitem{Effective}
E. Witten,
{\it Five-brane effective action in M-theory},
J. Geom. Phys. {\bf 22} (1997) 103-133, \newline
[{\tt arXiv:hep-th/9610234}].

\bibitem{Among}
E. Witten,
{\it Duality relations among topological effects in string theory},
J. High Energy Phys. {\bf 0005} (2000) 031,
[{\tt arXiv:hep-th/9912086}].

\bibitem{Zag}
D. Zagier, 
{\it Note on the Landweber-Stong elliptic genus},
in Elliptic curves and modular forms in algebraic topology (Princeton, NJ, 1986), 216--224, 
Lecture Notes in Math., 1326, Springer, Berlin, 1988. 



\end{thebibliography}
\end{document}